\documentclass[aps,pra]{revtex4}
\usepackage{graphics}

\makeatletter

\providecommand{\LyX}{L\kern-.1667em\lower.25em\hbox{Y}\kern-.125emX\@}

\usepackage{graphics}

\makeatletter

\usepackage{amsfonts}

\makeatother

\makeatother
\begin{document}

\title{Coherent Tunneling of Atoms from Bose-condensed Gases at Finite Temperatures}

\author{David L. Luxat}

\email{luxat@physics.utoronto.ca}

\author{Allan Griffin}

\email{griffin@physics.utoronto.ca}

\affiliation{Department of Physics, University of Toronto, Toronto, Ontario, Canada
M5S 1A7}

\begin{abstract}
Tunneling of atoms between two trapped Bose-condensed gases at finite
temperatures is explored using a many-body linear response tunneling
formalism similar to that used in superconductors. To lowest order,
the tunneling currents can be expressed quite generally in terms of
the single-particle Green's functions of the isolated Bose gases.
A coherent first-order tunneling Josephson current between two atomic
Bose-condensates is found, in addition to coherent and dissipative
contributions from second-order condensate-noncondensate and noncondensate-noncondensate
tunneling. Our work is a generalization of Meier and Zwerger, who
recently treated tunneling between uniform atomic Bose gases. We apply
our formalism to the analysis of an out-coupling experiment induced
by light wave fields, using a simple Bogoliubov-Popov quasiparticle
approximation for the trapped Bose gas. For tunneling into the vacuum,
we recover the results of Japha, Choi, Burnett and Band, who recently
pointed out the usefulness of studying the spectrum of out-coupled
atoms. In particular, we show that the small tunneling current of
noncondensate atoms from a trapped Bose gas has a broad spectrum of
energies, with a characteristic structure associated with the Bogoliubov
quasiparticle \( u^{2} \) and \( v^{2} \) amplitudes. 
\end{abstract}
\maketitle

\section{Introduction}

The ability to coherently transfer atoms out of a Bose-Einstein condensate
(BEC) in a trapped atomic gas is of fundamental importance in the
creation of coherent matter wave generators, often called {}``atom
lasers''. Theoretical interest has developed toward characterizing
the out-coupling stage of such an atom laser, with the hope of developing
efficiency criteria for this process. All this work has been done
in the zero temperature limit, using either a non-interacting gas
coupled to a laser mode \cite{moy-savage} or the modified Gross-Pitaevskii
dynamics of the condensate order parameter \cite{generalized-gp}.
A complete understanding of the out-coupling process necessarily must
take into account the presence of the thermal cloud composed of noncondensate
atoms. Recently, the role that the noncondensate atoms play in the
out-coupling process has been discussed by Japha, Choi, Burnett and
Band \cite{burnett-1,burnett-2}. In their work, a {}``pair-breaking''
contribution was obtained, in which a correlated pair of atoms take
part in the output process, with one output atom appearing simultaneously
with the creation of an excitation in the trapped Bose gas.

Out-coupling experiments are only a small class of experiments that
can be performed in which a BEC is coupled to another gas of atoms
(that may or may not be Bose-condensed). The MIT group \cite{structure-factor-phonons-exp,structure-factor-phonons-houches-exp}
has performed work in which a BEC is scattered from an optical Bragg
grating. They show that the excitation spectrum of the scattered atoms
gives a direct experimental measure of the dynamic structure factor
of the initial Bose-condensate. Since the detailed form of the dynamic
structure factor \( S(\mathbf{q},\omega ) \) is dependent upon the
many-body correlations of the Bose-condensate, this kind of experiment
provides a means of probing the unique correlations induced by a Bose-condensate
\cite{timmermans-tommasini-structure-factor-lda,trento-dynamic-structure-factor,otago-bragg-structure-factor}.

Despite the large amount of work that has been performed with regard
to the coupling of Bose-condensed gases with another (possibly Bose-condensed)
gas, there still does not exist in the literature a clean formulation
of such a coupling. Even though Refs.\( \:  \)\cite{burnett-1,burnett-2}
have sparked interest in this issue, the formalism used in these papers
does not show in a transparent way which correlations present in a
Bose-condensed gas can be probed by such experiments. Moreover, the
case in which both trapped gases are Bose-condensed was not considered.

In this paper, we set up a general linear response theory approach
for treating the coupling of two Bose gases at finite temperatures,
which gives a formally exact expression for the tunneling current.
The general expressions we derive --- within linear response theory
--- are independent of the specific approximation used for the many-body
states of the two uncoupled, trapped Bose gases. In this sense, our
results are the analogue of showing how the inelastic neutron scattering
cross-section is related to the dynamic structure factor \cite{griffins-book,trento-dynamic-structure-factor,otago-bragg-structure-factor}
and the general theory of tunneling in superconductors \cite{mahan,barone}.
The work closest to ours is by Meier and Zwerger \cite{finite-T-josephson} \nocite{pra-meier-zwerger},
who applied a similar formalism in discussing coherent Josephson tunneling
between two uniform Bose-condensed gases. In Sections \ref{section:linear response theory-general}
and \ref{Section: Tunneling currents in terms of correlations functions},
we give a detailed account of the many-body theory needed to find
formal expressions for the tunneling currents in terms of single-atom
correlation functions.

We apply our theoretical formalism to a class of experiments involving
Raman scattering of atoms between trapped Bose gases. We calculate
the tunneling current expression at finite temperatures using the
Bogoliubov-Popov approximation for the single-particle excitations
in the trapped Bose gases \cite{griffin-gapless}. We discuss coherent
tunneling of atoms between two Bose-condensed gases, as well as the
out-coupling of atoms from a BEC in a trap to the vacuum (i.e.,~to
free particle states). In this latter case, as expected, we show that
our more general formalism reproduces the results of Refs.\( \:  \)\cite{burnett-1,burnett-2}
in the appropriate limit. In Section \ref{sec:LDA}, we give an explicit
calculation of the out-coupling current based on treating the noncondensed
atoms within a simple local density approximation (LDA), such as used
in Refs.~\cite{timmermans-tommasini-structure-factor-lda,trento-dynamic-structure-factor}.

In essence, we show that besides the dominant contribution from tunneling
of atoms out of the condensate, one finds a much smaller contribution
which arises from the atoms tunneling out of the noncondensate. This
latter contribution, however, is of special interest since it is directly
related to the single-particle correlation function of the noncondensate
atoms. One component has weight proportional to the Bogoliubov amplitude
\( u_{E}^{2}N^{0}(E) \), where \( N^{0}(E) \) is the Bose-distribution
function for an excitation of energy \( E \), and increases with
temperature. The other, more interesting, component has weight proportional
to the Bogoliubov amplitude \( v_{E}^{2}(1+N^{0}(E)) \). This latter
process is formally related to the {}``negative energy pole'' of
the Bogoliubov spectral density. This so-called {}``pair-breaking''
term \cite{burnett-1,burnett-2} is of interest as a unique signature
of the unusual correlations induced by a Bose condensate. It corresponds
to the out-coupling of an atom and the simultaneous creation of a
Bogoliubov excitation in the trapped gas. This contribution already
is present at zero temperature and, in contrast to the tunneling current
from the condensate, contributes to an output tunneling current of
atoms with a wide spectrum of energies. As the temperature increases,
the weight of the \( v_{E}^{2} \) term decreases while the \( u_{E}^{2} \)
term starts to build up (we recall that as the condensate density
goes to zero, \( u_{E}^{2}\rightarrow 1 \) and \( v_{E}^{2}\rightarrow 0 \)).
The analogue of this pair-breaking process was discussed earlier in
the context of a Bose-condensed gas of excitons \cite{svg}.

In contrast to our work, a proposal has recently been put forward
\cite{brunello} to measure these Bogoliubov amplitudes using an optical
Bragg scattering technique. In this approach, a two-photon Bragg scattering
beam is used to excite quasiparticles out of the condensate and then
a second Bragg beam is used to probe the altered momentum distribution
of the sample. Recent work at MIT has beautifully verified these predictions
\cite{vogels-measuring-Bogoliubov}. Unlike this situation, in the
Raman scattering experiment the energy absorbed by the atoms imparts
a momentum kick in addition to causing a change of hyperfine state,
allowing the scattered atom to no longer be trapped.

Our results give another fairly direct method of measuring the Bogoliubov
quasiparticle amplitudes. Furthermore, since the weight of the out-coupling
current is proportional to the Bogoliubov amplitude \( v^{2} \),
it is significantly amplified for larger values of the s-wave scattering
length \( a \) such as can be achieved by working close to a Feshbach
resonance. We hope our work further stimulates experimental interest
in this kind of out-coupling experiment, even though the current from
the noncondensate atoms is quite small compared to that from the condensate.

\section{Linear response theory: general}

\label{section:linear response theory-general}Our goal is to calculate
the current of atoms tunneling between two atomic gases coupled through
a weak external field, which can be treated as a time-dependent perturbation
to a time-independent many-body Hamiltonian using linear response
theory (see, for example, Refs.\( \:  \)\cite{fetter-walecka,mahan}).
The total Hamiltonian in the Schr\"{o}dinger representation is\begin{equation}
\label{full hamiltonian}
\hat{H}(t)=\hat{H}_{0}+\varepsilon _{1}\hat{N}_{1}+\varepsilon _{2}\hat{N}_{2}+\hat{V}(t),
\end{equation}
 where \( \hat{H}_{0} \) is the Hamiltonian for the many-body systems,
\( \varepsilon _{1} \) and \( \varepsilon _{2} \) are the atomic
hyperfine level energies of an atom in gas 1 and gas 2 respectively,
\( \hat{N}_{1(2)} \) is the atomic number operator for gas 1(2),
given by\begin{equation}
\label{atomic number operator}
\hat{N}_{1(2)}=\int \mathrm{d}\mathbf{r}\hat{\psi }_{1(2)}^{\dagger }(\mathbf{r})\hat{\psi }_{1(2)}(\mathbf{r}).
\end{equation}
 The external probe field \( \hat{V}(t) \) is explicitly time-dependent.
We assume that the two many-body systems do not interact with one
another in the absence of \( \hat{V} \), so that \( \hat{H}_{0}=\hat{H}_{10}+\hat{H}_{20} \).
This amounts to assuming that the two gases do not overlap, \emph{i.e.~}the
coupling term of the kind \( \hat{\psi }_{1}^{\dagger }\hat{\psi }_{2}^{\dagger }\hat{\psi }_{2}\hat{\psi }_{1} \)
can be neglected. In the absence of \( \hat{V}(t) \), the number
of atoms in the two systems is fixed by their chemical potentials
\( \mu _{1} \) and \( \mu _{2} \).

The tunneling perturbation \( \hat{V}(t) \) couples the two many-body
systems together by introducing a mechanism by which an atom can tunnel
between the two systems. This perturbation has terms which create
an atom in one system while destroying an atom in the other system,
and vice-versa. An example of the kind of experiment described by
the tunneling perturbation \( \hat{V}(t) \) is the out-coupling phenomenon
considered in Refs.~\cite{burnett-1,burnett-2}. In this experiment,
an external electromagnetic (EM) field causes transitions between
two atomic hyperfine levels, one of which is not trapped. We label
the quantum states corresponding to these hyperfine levels by \( \left| 1\right\rangle  \)
and \( \left| 2\right\rangle  \). Assuming that hyperfine level \( \left| 1\right\rangle  \)
is lower in energy than hyperfine level \( \left| 2\right\rangle  \),
energy will be absorbed from (emitted to) the external field facilitating
the transition \( \left| 1\right\rangle \rightarrow \left| 2\right\rangle  \)
(\( \left| 2\right\rangle \rightarrow \left| 1\right\rangle  \)).
In more general experiments, however, the transition can take place
between two atomic hyperfine levels that are both trapped. An example
of such a process is a stimulated Raman transition between the magnetic
hyperfine levels \( \left| 1\right\rangle =\left| F=1,m_{F}=-1\right\rangle  \)
and \( \left| 2\right\rangle =\left| F=1,m_{F}=1\right\rangle  \)
(for a detailed discussion, see Ref.\ \cite{raman-output-coupler}).
We will return to the specifics of a BEC coupling experiment in Section
\ref{Section: Coupling between two trapped gases} when we calculate
the tunneling current between two trapped Bose-condensed gases.

For the moment it suffices to write down the semiclassical form of
the tunneling perturbation \( \hat{V} \) arising from the electric-dipole
and rotating wave approximations \cite{structure-factor-phonons-houches-exp,raman-output-coupler},
\begin{equation}
\label{V}
\hat{V}(t)=\mathrm{e}^{\eta t}\int \mathrm{d}\mathbf{r}\bigg (\gamma \mathrm{e}^{i(\mathbf{q}\cdot \mathbf{r}-\omega t)}\hat{\psi }_{1}(\mathbf{r})\hat{\psi }^{\dagger }_{2}(\mathbf{r})+\mathrm{h}.\mathrm{c}.\bigg ).
\end{equation}
 Here \( \omega  \) is the effective energy transfer from the laser
fields, \( \mathbf{q} \) is the momentum transfer from the laser
fields, and \( \gamma  \) (taken to be real) is the interaction strength
that is associated with the Rabi frequency of the transition between
atomic states. We have also introduced a factor \( \mathrm{e}^{\eta t} \)
that models the adiabatic switching on of the interaction at \( t=-\infty  \).
In our final results, we take the limit \( \eta \rightarrow 0^{+} \).
We use units in which \( \hbar =1 \) and \( k_{B}=1 \), unless otherwise
specified.

We next outline the formalism for calculating the tunneling current
between two Bose gases (resulting from the action of \( \hat{V} \)).
The state of the combined system evolves in time according to \( \left| \psi (t)\right\rangle =\hat{U}(t)\left| \psi (-\infty )\right\rangle  \),
where the evolution operator \( \hat{U}(t) \) is unitary and satisfies
the equation of motion\begin{eqnarray}
i\frac{\partial \hat{U}(t)}{\partial t} & = & \hat{U}(t)\hat{H}(t).\label{equation of motion for time-evolution operator} 
\end{eqnarray}
 This equation of motion can be formally solved to give \( \hat{U}(t)=\mathrm{T}\exp \bigg (-i\int _{-\infty }^{t}\mathrm{d}t^{\prime }\hat{H}(t^{\prime })\bigg ) \),
where \( \mathrm{T} \) denotes the time-ordering operator. In the
Heisenberg picture, operators are written in terms of the corresponding
Schr\"{o}dinger picture operators as \( \hat{A}_{H}(t)\equiv \hat{U}^{\dagger }(t)\hat{A}\hat{U}(t) \),
while the state of the system is given by the initial state of the
system at \( t=-\infty  \), namely \( \left| \psi \right\rangle _{H}\equiv \left| \psi (-\infty )\right\rangle  \).
We use the subscript \( H \) to denote an operator or state in the
Heisenberg picture with respect to the full Hamiltonian \( \hat{H} \).
The Heisenberg equation of motion for any operator \( \hat{A} \)
is then given as\begin{equation}
\label{Heisenberg eom for operator A}
\frac{\partial \hat{A}_{H}(t)}{\partial t}=i\left[ \hat{H}_{H}(t),\hat{A}_{H}(t)\right] .
\end{equation}

We define the atomic current \emph{from} gas 1 \emph{to} gas 2 in
terms of the thermal equilibrium ensemble average of \( \dot{\hat{N}}_{1H} \)\begin{equation}
\label{current definition}
I(t)\equiv -\left\langle \dot{\hat{N}}_{1H}(t)\right\rangle .
\end{equation}
 We have introduced a negative sign in this expression for the atomic
current so that the current flow from gas 1 to gas 2 is positive.
The ensemble average in (\ref{current definition}) is given by \begin{eqnarray}
\left\langle \dot{\hat{N}}_{1H}(t)\right\rangle  & = & \mathrm{Tr}\left\{ \hat{\rho }_{eq}\dot{\hat{N}}_{1H}(t)\right\} \nonumber \\
 & = & i\mathrm{Tr}\left\{ \hat{\rho }_{eq}\hat{U}^{\dagger }(t)\left[ \hat{H}(t),\hat{N}_{1}\right] \hat{U}(t)\right\} ,\label{ensemble average for N1dot 1} 
\end{eqnarray}
 where the initial equilibrium density operator \( \hat{\rho }_{eq} \)
describes the state of the system at \( t=-\infty  \) before the
coupling \( \hat{V} \) is turned on. Since \( \left[ \hat{H}(t),\hat{N}_{1}\right] =\left[ \hat{V}(t),\hat{N}_{1}\right]  \),
the average in (\ref{ensemble average for N1dot 1}) takes the form\begin{equation}
\label{ensemble average for N1dot 2}
\left\langle \dot{\hat{N}}_{1H}(t)\right\rangle =i\mathrm{Tr}\left\{ \hat{\rho }_{eq}\hat{U}^{\dagger }(t)\left[ \hat{V}(t),\hat{N}_{1}\right] \hat{U}(t)\right\} .
\end{equation}
 We now turn to the task of evaluating this ensemble average within
linear response theory \cite{fetter-walecka,mahan}.

We first write the average (\ref{ensemble average for N1dot 2}) in
an interaction picture, in which \( \hat{V} \) is taken as a perturbation
to the many-body Hamiltonian \( \hat{H}_{0}^{\prime }\equiv \hat{H}_{0}+\varepsilon _{1}\hat{N}_{1}+\varepsilon _{2}\hat{N}_{2} \)
of the uncoupled gases. Later we will introduce an interaction picture
involving \( \hat{K}_{0}\equiv \hat{H}_{0}^{\prime }-\mu _{1}\hat{N}_{1}-\mu _{2}\hat{N}_{2} \).
With respect to \( \hat{H}_{0}^{\prime } \), we define the interaction
representation as \begin{equation}
\label{H0prime interaction representation operator defined}
\hat{A}_{H_{0}^{\prime }}(t)\equiv \mathrm{e}^{-i\hat{H}_{0}^{\prime }t}\hat{A}\mathrm{e}^{i\hat{H}_{0}^{\prime }t}=\mathrm{e}^{-i\hat{H}_{0}^{\prime }t}\hat{U}(t)\hat{A}_{H}(t)\hat{U}^{\dagger }(t)\mathrm{e}^{i\hat{H}_{0}^{\prime }t}.
\end{equation}
 In this interaction picture, we also define the time-dependent density
operator as \begin{equation}
\label{density matrix in interaction picture}
\hat{\rho }_{H_{0}^{\prime }}(t)\equiv \mathrm{e}^{i\hat{H}_{0}^{\prime }t}\hat{U}(t)\hat{\rho }_{eq}\hat{U}^{\dagger }(t)\mathrm{e}^{-i\hat{H}_{0}^{\prime }t}.
\end{equation}
 Using these definitions, we now can write (\ref{ensemble average for N1dot 2})
as \begin{equation}
\label{ensemble average for N1dot 3}
\left\langle \dot{\hat{N}}_{1H}(t)\right\rangle =i\mathrm{Tr}\left\{ \hat{\rho }_{H_{0}^{\prime }}(t)\mathrm{e}^{i\hat{H}_{0}^{\prime }t}\left[ \hat{V}(t),\hat{N}_{1}\right] \mathrm{e}^{-i\hat{H}_{0}^{\prime }t}\right\} .
\end{equation}
 Evaluating the commutator \( \left[ \hat{V}(t),\hat{N}_{1}\right]  \)
in (\ref{ensemble average for N1dot 3}), we obtain\begin{equation}
\label{ensemble average for N1dot 4}
\left\langle \dot{\hat{N}}_{1H}(t)\right\rangle =i\int \mathrm{d}\mathbf{r}\Biggl [\gamma \mathrm{e}^{i(\mathbf{q}\cdot \mathbf{r}-\omega t)}\mathrm{Tr}\left\{ \hat{\rho }_{H_{0}^{\prime }}(t)\hat{\psi }_{1H_{0}^{\prime }}(\mathbf{r},t)\hat{\psi }_{2H_{0}^{\prime }}(\mathbf{r},t)\right\} -\mathrm{c}.\mathrm{c}.\Biggr ].
\end{equation}

The density operator in the interaction picture defined in (\ref{density matrix in interaction picture})
satisfies the Liouville equation \begin{equation}
\label{Liouville equation in interaction picture}
\frac{\partial \hat{\rho }_{H_{0}^{\prime }}(t)}{\partial t}=i\left[ \hat{\rho }_{H_{0}^{\prime }}(t),\hat{V}_{H_{0}^{\prime }}(t)\right] ,
\end{equation}
 which has a formal solution\begin{equation}
\label{Liouville equation in integral form}
\hat{\rho }_{H_{0}^{\prime }}(t)=\hat{\rho }(-\infty )+i\int _{-\infty }^{t}\mathrm{d}t^{\prime }\left[ \hat{\rho }_{H_{0}^{\prime }}(t^{\prime }),\hat{V}_{H_{0}^{\prime }}(t^{\prime })\right] .
\end{equation}
 Iterating (\ref{Liouville equation in integral form}) to first order
in \( \hat{V} \), we obtain \begin{equation}
\label{linear response theory approximation to density matrix in interaction picture}
\hat{\rho }_{H_{0}^{\prime }}(t)=\hat{\rho }_{eq}+\delta \hat{\rho }_{H_{0}^{\prime }}(t).
\end{equation}
 Here\begin{equation}
\label{linear response density matrix in interaction picture}
\delta \hat{\rho }_{H_{0}^{\prime }}(t)\equiv i\int _{-\infty }^{t}\mathrm{d}t^{\prime }\left[ \hat{\rho }_{eq},\hat{V}_{H_{0}^{\prime }}(t^{\prime })\right] 
\end{equation}
 is the perturbation to the initial equilibrium density operator \( \hat{\rho }_{eq} \)
to first order in \( \hat{V} \), the linear response approximation
to the density matrix.

Within this first-order approximation for \( \hat{\rho }_{H_{0}^{\prime }}(t) \)
the ensemble average of an operator \( \hat{A} \) is given by \( \left\langle \hat{A}_{H}(t)\right\rangle =\left\langle \hat{A}_{H_{0}^{\prime }}(t)\right\rangle _{eq}+\delta \left\langle \hat{A}_{H_{0}^{\prime }}(t)\right\rangle , \)
where \begin{eqnarray}
\left\langle \hat{A}_{H_{0}^{\prime }}(t)\right\rangle _{eq} & \equiv  & \mathrm{Tr}\left\{ \hat{\rho }_{eq}\hat{A}_{H_{0}^{\prime }}(t)\right\} \label{equlibrium ensemble average} \\
\delta \left\langle \hat{A}_{H_{0}^{\prime }}(t)\right\rangle  & \equiv  & \mathrm{Tr}\left\{ \delta \hat{\rho }_{H_{0}^{\prime }}(t)\hat{A}_{H_{0}^{\prime }}(t)\right\} .\label{linear response ensemble average} 
\end{eqnarray}
 The atomic current given by (\ref{current definition}) can thus
be written in linear response theory as\begin{equation}
\label{total current - linear response}
I(t)=I_{eq}(t)+\delta I(t),
\end{equation}
 where \( I_{eq} \) is the equilibrium current and \( \delta I \)
is the linear response current. Substituting our approximation for
\( \hat{\rho }_{H_{0}^{\prime }}(t) \) of (\ref{linear response theory approximation to density matrix in interaction picture})
together with (\ref{V}) for \( \hat{V} \) into (\ref{ensemble average for N1dot 4}),
we obtain, after some algebra, explicit expressions for the equilibrium
current \begin{equation}
\label{equilibrium current 1}
I_{eq}(t)=2\mathrm{Im}\int \mathrm{d}\mathbf{r}\gamma \mathrm{e}^{i(\mathbf{q}\cdot \mathbf{r}-\omega t)}\left\langle \hat{\psi }_{1H_{0}^{\prime }}(\mathbf{r},t)\hat{\psi }^{\dagger }_{2H_{0}^{\prime }}(\mathbf{r},t)\right\rangle _{eq}
\end{equation}
 and the linear response current \begin{eqnarray}
\delta I(t) & = & 2\mathrm{Re}\int \mathrm{d}\mathbf{r}\int \mathrm{d}\mathbf{r}^{\prime }\int _{-\infty }^{t}\mathrm{d}t^{\prime }\mathrm{e}^{\eta (t+t^{\prime })}\nonumber \\
 &  & \times \Biggl \{\gamma ^{2}\mathrm{e}^{i[\mathbf{q}\cdot (\mathbf{r}-\mathbf{r}^{\prime })-\omega (t-t^{'})]}\left\langle \left[ \hat{\psi }^{\dagger }_{1H_{0}^{\prime }}(\mathbf{r}^{\prime },t^{\prime })\hat{\psi }_{2H_{0}^{\prime }}(\mathbf{r}^{\prime },t^{\prime }),\hat{\psi }_{1H_{0}^{\prime }}(\mathbf{r},t)\hat{\psi }^{\dagger }_{2H_{0}^{\prime }}(\mathbf{r},t)\right] \right\rangle _{eq}\nonumber \\
 &  & +\gamma ^{2}\mathrm{e}^{i[\mathbf{q}\cdot (\mathbf{r}+\mathbf{r}^{\prime })-\omega (t+t^{'})]}\left\langle \left[ \hat{\psi }_{1H_{0}^{\prime }}(\mathbf{r}^{\prime },t^{\prime })\hat{\psi }^{\dagger }_{2H_{0}^{\prime }}(\mathbf{r}^{\prime },t^{\prime }),\hat{\psi }_{1H_{0}^{\prime }}(\mathbf{r},t)\hat{\psi }^{\dagger }_{2H_{0}^{\prime }}(\mathbf{r},t)\right] \right\rangle _{eq}\Biggr \}.\label{linear response current 1} 
\end{eqnarray}
 The expressions in (\ref{equilibrium current 1}) and (\ref{linear response current 1})
are the basis of the rest of this paper. One sees that there are two
different contributions to the linear response current in (\ref{linear response current 1}),
corresponding to phases depending on the times through \( t-t^{\prime } \)
or \( t+t^{\prime } \). The former will be seen shortly to correspond
to a \emph{normal} current of single atoms between the two gases,
while the latter corresponds to an \emph{anomalous} current of atoms
between the two gases.

We next want to express the ensemble averages in (\ref{equilibrium current 1})
and (\ref{linear response current 1}) in terms of thermal Green's
functions given by standard techniques in finite temperature many-body
theory \cite{mahan,fetter-walecka}. To this end, we first express
the time-dependent field operators in (\ref{equilibrium current 1})
and (\ref{linear response current 1}) in the usual {}``Heisenberg
picture'' used in many-body theory, namely the interaction picture
in which the free Hamiltonian is taken to be the many-body {}``grand
canonical Hamiltonian'' \begin{equation}
\label{K0}
\hat{K}_{0}\equiv \hat{H}_{0}^{\prime }-\mu _{1}\hat{N}_{1}-\mu _{2}\hat{N}_{2}.
\end{equation}
 Using the identity \( \mathrm{e}^{\lambda \hat{N}_{i}}\hat{\psi }_{i}(\mathbf{r})\mathrm{e}^{-\lambda \hat{N}_{i}}=\mathrm{e}^{-\lambda }\hat{\psi }_{i}(\mathbf{r}) \),
it is easily verified that \
\begin{equation}
\label{relation between destruction operators in H0prime and K0 interaction pictures}
\hat{\psi }_{iH_{0}^{\prime }}(\mathbf{r},t)=\mathrm{e}^{-i(\epsilon _{i}+\mu _{i})t}\hat{\psi }_{iK_{0}}(\mathbf{r},t),
\end{equation}
 where we define \( \hat{\psi }_{iK_{0}}(\mathbf{r},t)\equiv \mathrm{e}^{i\hat{K}_{0}t}\hat{\psi }_{i}(\mathbf{r})\mathrm{e}^{-i\hat{K}_{0}t} \).
Using (\ref{relation between destruction operators in H0prime and K0 interaction pictures}),
it is easy to rewrite (\ref{linear response current 1}) in terms
of the \( \hat{\psi }_{iK_{0}} \) operators. We also note that the
grand canonical ensemble average of a product of operators of the
two gases factorizes \
\begin{equation}
\label{ensemble average decomposition}
\left\langle \hat{A}_{1K_{0}}\hat{A}_{2K_{0}}\right\rangle _{eq}\rightarrow \left\langle \hat{A}_{1K_{0}}\right\rangle _{1eq}\left\langle \hat{A}_{2K_{0}}\right\rangle _{2eq},
\end{equation}
 where \( \left\langle \cdot \cdot \cdot \right\rangle _{1eq}\equiv \mathrm{Tr}\left\{ \hat{\rho }_{1eq}(\cdot \cdot \cdot )\right\}  \).

Below the transition temperature of a Bose gas, the new phase of matter
is described by a symmetry-broken state in which the equilibrium density
operator does \emph{not} commute with the number operator: \( \left[ \hat{\rho }_{eq},\hat{N}\right] \neq 0 \)
(the state breaks the system's \( \mathrm{U}(1) \) symmetry). Thus,
the equilibrium ensemble averages are truly over the \( \eta  \)-ensemble
introduced by Bogoliubov \cite{bogoliubov-lecture-vol2}, which is
symbolically defined as \
\begin{equation}
\label{eta ensemble broken symmetry average}
\left\langle \hat{A}_{1K_{0}}\right\rangle _{eq}\equiv \lim _{\eta \rightarrow 0}\mathrm{Tr}\left\{ \hat{\rho }_{1eq}^{\eta }\hat{A}_{1K_{0}}\right\} ,
\end{equation}
 where \
\begin{equation}
\label{broken symmetry grand canonical density operator}
\hat{\rho }^{\eta }_{1eq}\equiv \frac{\mathrm{e}^{-\beta _{1}(\hat{H}_{10}+\hat{H}_{1sb}-\mu _{1}\hat{N}_{1})}}{Z^{\eta }_{1}}
\end{equation}
 is the symmetry-breaking grand canonical equilibrium density operator.
The symmetry-breaking Hamiltonian is defined as \cite{bogoliubov-lecture-vol2} \
\begin{equation}
\label{broken symmetry Hamiltonian}
\hat{H}_{isb}\equiv \int \mathrm{d}\mathbf{r}\left( \eta ^{*}_{i}(\mathbf{r})\hat{\psi }_{i}(\mathbf{r})+\eta _{i}(\mathbf{r})\hat{\psi }^{\dagger }_{i}(\mathbf{r})\right) .
\end{equation}
 \( Z^{\eta }_{1}\equiv \mathrm{Tr}\bigg \{\mathrm{e}^{-\beta _{1}(\hat{H}_{10}+\hat{H}_{1sb}-\mu _{1}\hat{N}_{1})}\bigg \} \)
is the broken symmetry grand canonical ensemble partition function.
We will use the notation \( \left\langle \cdots \right\rangle _{eq} \)
to denote equilibrium ensemble averages, with the understanding that
they are over a broken symmetry state. An equivalent way of stating
this is that atomic field operators can be decomposed into two parts:
\
\begin{equation}
\label{order+fluctuation-field}
\hat{\psi }_{1K_{0}}(\mathbf{r},t)=\Phi _{10}(\mathbf{r})+\tilde{\psi }_{1K_{0}}(\mathbf{r},t),
\end{equation}
 where \
\begin{equation}
\label{order-parameter}
\Phi _{0}(\mathbf{r})\equiv \left\langle \hat{\psi }_{K_{0}}(\mathbf{r},t)\right\rangle _{eq}\equiv \sqrt{n_{c0}(\mathbf{r})}\mathrm{e}^{i\theta _{c0}(\mathbf{r})}
\end{equation}
 is the symmetry-breaking order parameter field and \( \tilde{\psi } \)
is the fluctuation field operator associated with the noncondensate
atoms. In (\ref{order-parameter}), \( n_{c0} \) is the local condensate
density while \( \theta _{c0} \) is the condensate phase. In our
formalism based on linear response from thermal equilibrium, the Bose
order parameter in (\ref{order+fluctuation-field}) is that appropriate
to thermal equilibrium. A phase \( \theta _{c0}(\mathbf{r}) \) would
arise if we were dealing, for example, with a condensate with a vortex.
We also note that the usual time-dependence of the Bose-order parameter
(\( \mathrm{e}^{-i\mu t} \)) is absent because we are working in
the \( \hat{K}_{0} \) interaction representation.

We do not need to assume that the inverse temperature is the same
for both gases. Indeed, because the two gases are assumed to be uncoupled
in the absence of the perturbation \( \hat{V} \), no mechanism exists
to force both gases to reach a common thermodynamic equilibrium. Therefore,
the equilibrium ensemble averages in (\ref{equilibrium current 1}),
(\ref{linear response current 1}) and (\ref{ensemble average decomposition})
may be viewed as being over two independent gases.

The equilibrium tunneling current in (\ref{equilibrium current 1})
is first order in the perturbation \( \hat{V} \), whereas the linear
response current in (\ref{linear response current 1}) is second order
in \( \hat{V} \). Using (\ref{relation between destruction operators in H0prime and K0 interaction pictures}),
(\ref{equilibrium current 1}) can be written as \begin{equation}
\label{equilibrium current 2}
I_{eq}(t)=2\mathrm{Im}\int \mathrm{d}\mathbf{r}\gamma \mathrm{e}^{i\mathbf{q}\cdot \mathbf{r}}\mathrm{e}^{-i(\omega -\Delta \varepsilon -\Delta \mu )t}\left\langle \hat{\psi }_{1K_{0}}(\mathbf{r},t)\right\rangle _{1eq}\left\langle \hat{\psi }^{\dagger }_{2K_{0}}(\mathbf{r},t)\right\rangle _{2eq},
\end{equation}
 where \( \Delta \varepsilon \equiv \varepsilon _{2}-\varepsilon _{1} \),
\( \Delta \mu \equiv \mu _{2}-\mu _{1} \). Introducing (\ref{order-parameter})
into (\ref{equilibrium current 2}), the equilibrium current is reduced
to\begin{equation}
\label{equilibrium current 3}
I_{eq}(t)=-2\gamma \int \mathrm{d}\mathbf{r}\sqrt{n_{c10}(\mathbf{r})n_{c20}(\mathbf{r})}\sin \biggl [(\omega -\Delta \varepsilon -\Delta \mu )t-\mathbf{q}\cdot \mathbf{r}+\theta _{c20}(\mathbf{r})-\theta _{c10}(\mathbf{r})\biggr ].
\end{equation}
 The magnitude of this equilibrium current is seen to be proportional
to the overlap of the two condensate density profiles and thus serves
as a probe of the condensate surface structure.

In a non-Bose condensed gas, the ensemble averages \( \left\langle \hat{\psi }_{1K_{0}}(\mathbf{r},t)\right\rangle _{1eq} \)
and \( \left\langle \hat{\psi }_{2K_{0}}(\mathbf{r},t)\right\rangle _{2eq} \)
in (\ref{equilibrium current 2}) vanish. As noted in (\ref{order-parameter})
and the discussion preceding it, these ensemble averages do \emph{not}
vanish in a Bose-condensed system. Thus, the equilibrium current (\ref{equilibrium current 3})
is a direct manifestation of Bose broken-symmetry. This expression
for the equilibrium current is a more general form of the \emph{linearized}
{}``internal'' Josephson current \cite{jamies-josephson,leggett-review-2000}
for Bose-condensed gases. Our expression (\ref{equilibrium current 3})
differs from previous results in several ways. Firstly, there is the
phase associated with the momentum \( \mathbf{q} \) transferred by
the two laser beams. This will introduce momentum kicks (Bragg scattering)
to the atoms in the two condensates \cite{structure-factor-phonons-houches-exp}.
Secondly, we have allowed for the case of a spatially varying condensate
phase difference, \( \Delta \theta (\mathbf{r})\equiv \theta _{c20}(\mathbf{r})-\theta _{c10}(\mathbf{r}) \),
such as would arise if one of the condensates had a vortex.

The quantity \( \omega -\Delta \varepsilon -\Delta \mu \) in (\ref{equilibrium current 3})
is analogous to the voltage in the standard discussion of the Josephson
current in superconductors \cite{mahan}. We will show later (see
(\ref{deltaIAa}) and (\ref{deltaIAb})) that this quantity also enters
into the phase of the anomalous component of the linear response current
in (\ref{linear response current 1}) as a second harmonic. We may
interpret the atomic energy level splitting of the atomic states \( \Delta \varepsilon  \)
in addition to the chemical potential difference \( \Delta \mu  \)
of the two Bose gases as a negative voltage impeding the atomic current
flow \emph{from} gas 1 \emph{to} gas 2. The energy difference between
the laser photons acts as a forward biased voltage serving to negate
this reverse bias. In the regime where \( \omega -\Delta \varepsilon -\Delta \mu \rightarrow 0 \),
we recover what is termed the d\@.c\@.~Josephson effect in superconducting
systems \cite{mahan}. In this regime, as with superconductors, a
non-zero difference between the phases of the two condensate order
parameters drives a d\@.c\@.~atomic current.

\section{Tunneling currents in terms of correlation functions}

\label{Section: Tunneling currents in terms of correlations functions}We
next turn to the linear response current in (\ref{linear response current 1}),
which is second order in the tunneling perturbation \( \hat{V} \).
We first want to write it in terms of thermal Green's functions by
introducing correlation functions of the atomic field operators. As
in the previous section, we rewrite our time-dependent interaction
picture field operators in terms of a Heisenberg picture defined with
respect to \( \hat{K}_{0} \), rather than \( \hat{H}_{0}^{\prime } \)
(see (\ref{relation between destruction operators in H0prime and K0 interaction pictures})).

We consider the first term in the expression of the linear response
current in (\ref{linear response current 1}). We refer to this term
as the normal current, for reasons which shall become clear. The normal
current is then given by (\( \Delta \varepsilon \equiv \varepsilon _{2}-\varepsilon _{1},\Delta \mu \equiv \mu _{2}-\mu _{1} \))
\begin{eqnarray}
\delta I_{N}(t) & \equiv  & 2\gamma ^{2}\, \mathrm{Re}\int _{-\infty }^{t}\mathrm{d}t^{\prime }\mathrm{e}^{\eta (t+t^{\prime })}\int \mathrm{d}\mathbf{r}^{\prime }\int \mathrm{d}\mathbf{r}^{\prime \prime }\mathrm{e}^{i\mathbf{q}\cdot (\mathbf{r}-\mathbf{r}^{\prime })}\mathrm{e}^{-i(\omega -\Delta \varepsilon -\Delta \mu )(t-t^{\prime })}\nonumber \\
 &  & \times \biggl [C^{K_{0}}_{\hat{\psi }^{\dagger }_{1}\hat{\psi }_{1}}(\mathbf{r}^{\prime },\mathbf{r},t^{\prime }-t)C^{K_{0}}_{\hat{\psi }_{2}\hat{\psi }_{2}^{\dagger }}(\mathbf{r}^{\prime },\mathbf{r},t^{\prime }-t)\nonumber \\
 &  & \qquad -C^{K_{0}}_{\hat{\psi }_{1}\hat{\psi }^{\dagger }_{1}}(\mathbf{r},\mathbf{r}^{\prime },t-t^{\prime })C^{K_{0}}_{\hat{\psi }_{2}^{\dagger }\hat{\psi }_{2}}(\mathbf{r},\mathbf{r}^{\prime },t-t^{\prime })\biggr ].\label{normal current definition} 
\end{eqnarray}
 We have evaluated the commutators in (\ref{linear response current 1})
making use of (\ref{ensemble average decomposition}) to write the
ensemble averages as a product of two-point correlation functions
such as \begin{equation}
\label{two-point correlation function definition}
C^{K_{0}}_{\hat{\psi }_{1}^{\dagger }\hat{\psi }_{1}}(\mathbf{r},\mathbf{r}^{\prime },t)\equiv \left\langle \hat{\psi }_{1K_{0}}^{\dagger }(\mathbf{r},t)\hat{\psi }_{1K_{0}}(\mathbf{r}^{\prime })\right\rangle _{1eq}.
\end{equation}
 In subsequent discussion, we will refer to \( \hat{\psi }_{K_{0}}(\mathbf{r},t) \)
as \( \hat{\psi }(\mathbf{r},t) \) for notational simplicity. The
normal current given in (\ref{normal current definition}) is independent
of the time \( t \). We make this more explicit by rewriting the
integral and noting that factors of \( \mathrm{e}^{\eta t} \) factor
can be neglected in the limit \( \eta \rightarrow 0^{+} \). Introducing
the (Heaviside) step function \( \theta  \), the normal current may
be written in the explicitly time-independent form \begin{eqnarray}
\delta I_{N} & = & 2\gamma ^{2}\, \mathrm{Re}\int _{-\infty }^{\infty }\mathrm{d}\tau \mathrm{e}^{-\eta \tau }\theta (\tau )\int \mathrm{d}\mathbf{r}\int \mathrm{d}\mathbf{r}^{\prime }\mathrm{e}^{i\mathbf{q}\cdot (\mathbf{r}-\mathbf{r}^{\prime })}\mathrm{e}^{-i(\omega -\Delta \varepsilon -\Delta \mu )\tau }\nonumber \\
 &  & \; \times \biggl [C_{\hat{\psi }^{\dagger }_{1}\hat{\psi }_{1}}(\mathbf{r}^{\prime },\mathbf{r},-\tau )C_{\hat{\psi }_{2}\hat{\psi }_{2}^{\dagger }}(\mathbf{r}^{\prime },\mathbf{r},-\tau )-C_{\hat{\psi }_{1}\hat{\psi }^{\dagger }_{1}}(\mathbf{r},\mathbf{r}^{\prime },\tau )C_{\hat{\psi }_{2}^{\dagger }\hat{\psi }_{2}}(\mathbf{r},\mathbf{r}^{\prime },\tau )\biggr ].\label{normal current 1} 
\end{eqnarray}
 We observe that, as expected for a current, the second term in (\ref{normal current 1})
breaks time-reversal symmetry; the current in the negative time direction
is the negative of that in the positive time direction.

The correlation functions appearing in (\ref{normal current 1}) may
now be written in terms of their Fourier transforms \begin{equation}
\label{correlation function fourier transform}
C_{\hat{\psi }^{\dagger }\hat{\psi }}(\mathbf{r},\mathbf{r}^{\prime },t)\equiv \int _{-\infty }^{\infty }\frac{\mathrm{d}\omega }{2\pi }\mathrm{e}^{-i\omega t}C_{\hat{\psi }^{\dagger }\hat{\psi }}(\mathbf{r},\mathbf{r}^{\prime },\omega ),
\end{equation}
 and hence (\ref{normal current 1}) becomes \begin{eqnarray}
\delta I_{N} & = & 2\gamma ^{2}\, \mathrm{Re}\int _{-\infty }^{\infty }\mathrm{d}\tau \int _{-\infty }^{\infty }\frac{\mathrm{d}\omega ^{\prime }}{2\pi }\int _{-\infty }^{\infty }\frac{\mathrm{d}\omega ^{\prime \prime }}{2\pi }\mathrm{e}^{-\eta \tau }\theta (\tau )\int \mathrm{d}\mathbf{r}\int \mathrm{d}\mathbf{r}^{\prime }\mathrm{e}^{i\mathbf{q}\cdot (\mathbf{r}-\mathbf{r}^{\prime })}\mathrm{e}^{i(\omega ^{\prime \prime }-\omega ^{\prime }-\omega +\Delta \varepsilon +\Delta \mu )\tau }\nonumber \\
 &  & \; \times \biggl [C_{\hat{\psi }^{\dagger }_{1}\hat{\psi }_{1}}(\mathbf{r}^{\prime },\mathbf{r},-\omega ^{\prime })C_{\hat{\psi }_{2}\hat{\psi }_{2}^{\dagger }}(\mathbf{r}^{\prime },\mathbf{r},\omega ^{\prime \prime })-C_{\hat{\psi }_{1}\hat{\psi }^{\dagger }_{1}}(\mathbf{r},\mathbf{r}^{\prime },\omega ^{\prime })C_{\hat{\psi }_{2}^{\dagger }\hat{\psi }_{2}}(\mathbf{r},\mathbf{r}^{\prime },-\omega ^{\prime \prime })\biggr ].\label{normal current 2} 
\end{eqnarray}
 The integral over the time \( \tau  \) in (\ref{normal current 2})
is trivially given by \begin{equation}
\label{time integral}
\int _{-\infty }^{\infty }\mathrm{d}\tau \theta (\tau )\mathrm{e}^{i(\omega ^{\prime \prime }-\omega ^{\prime }-\omega +\Delta \varepsilon +\Delta \mu +i\eta )\tau }=\frac{i}{\omega ^{\prime \prime }-\omega ^{\prime }-\omega +\Delta \varepsilon +\Delta \mu +i\eta }.
\end{equation}
 In the physical limit \( \eta \rightarrow 0^{+} \), we may use the
identity \begin{equation}
\label{complex analysis identity}
\lim _{\eta \rightarrow 0^{+}}\frac{1}{\omega +i\eta }=\mathrm{P}\left( \frac{1}{\omega }\right) -i\pi \delta (\omega ),
\end{equation}
 where \( \mathrm{P} \) denotes the principal value of an integral.
Therefore, the normal current (\ref{normal current 2}) may be written
in the form \begin{equation}
\label{normal current 3}
\delta I_{N}=\delta I_{Na}+\delta I_{Nb},
\end{equation}
 where we have introduced the two contributions \begin{eqnarray}
\delta I_{Na} & \equiv  & \gamma ^{2}\, \mathrm{Re}\int \mathrm{d}\mathbf{r}\int \mathrm{d}\mathbf{r}^{\prime }\mathrm{e}^{i\mathbf{q}\cdot (\mathbf{r}-\mathbf{r}^{\prime })}\int _{-\infty }^{\infty }\frac{\mathrm{d}\omega ^{\prime }}{2\pi }\nonumber \\
 &  & \qquad \biggl [C_{\hat{\psi }^{\dagger }_{1}\hat{\psi }_{1}}(\mathbf{r}^{\prime },\mathbf{r},-\omega ^{\prime })C_{\hat{\psi }_{2}\hat{\psi }_{2}^{\dagger }}\bigg (\mathbf{r}^{\prime },\mathbf{r},\omega ^{\prime }+\omega -\Delta \varepsilon -\Delta \mu \bigg )\nonumber \\
 &  & \qquad \qquad -C_{\hat{\psi }_{1}\hat{\psi }^{\dagger }_{1}}(\mathbf{r},\mathbf{r}^{\prime },\omega ^{\prime })C_{\hat{\psi }_{2}^{\dagger }\hat{\psi }_{2}}\bigg (\mathbf{r},\mathbf{r}^{\prime },-\omega ^{\prime }-\omega +\Delta \varepsilon +\Delta \mu \bigg )\biggr ]\label{deltaINa} 
\end{eqnarray}
 and \begin{eqnarray}
\delta I_{Nb} & \equiv  & -2\gamma ^{2}\, \mathrm{Im}\int \mathrm{d}\mathbf{r}\int \mathrm{d}\mathbf{r}^{\prime }\mathrm{e}^{i\mathbf{q}\cdot (\mathbf{r}-\mathbf{r}^{\prime })}\mathrm{P}\int _{-\infty }^{\infty }\frac{\mathrm{d}\omega ^{\prime }}{2\pi }\int _{-\infty }^{\infty }\frac{\mathrm{d}\omega ^{\prime \prime }}{2\pi }\nonumber \\
 &  & \frac{\biggl [C_{\hat{\psi }^{\dagger }_{1}\hat{\psi }_{1}}(\mathbf{r}^{\prime },\mathbf{r},-\omega ^{\prime })C_{\hat{\psi }_{2}\hat{\psi }_{2}^{\dagger }}(\mathbf{r}^{\prime },\mathbf{r},\omega ^{\prime \prime })-C_{\hat{\psi }_{1}\hat{\psi }^{\dagger }_{1}}(\mathbf{r},\mathbf{r}^{\prime },\omega ^{\prime })C_{\hat{\psi }_{2}^{\dagger }\hat{\psi }_{2}}(\mathbf{r},\mathbf{r}^{\prime },-\omega ^{\prime \prime })\biggr ]}{\omega ^{\prime \prime }-\omega ^{\prime }-\omega +\Delta \varepsilon +\Delta \mu }.\label{short time normal current} 
\end{eqnarray}

We next show how the correlation functions in (\ref{deltaINa}) and
(\ref{short time normal current}) may be converted to spectral densities,
which in turn can be directly calculated from time-ordered single-particle
Green's functions \cite{mahan}. Thus, the problem is reduced to the
application of well-known techniques from many-body theory to calculate
these Green's functions \cite{fetter-walecka,mahan}. We first rewrite
the correlation functions in terms of the order parameter fields and
the single-particle correlation functions of the noncondensate field
operators. The diagonal correlation functions above become \begin{equation}
\label{symmetry-breaking-correlation-1}
C_{\hat{\psi }^{\dagger }\hat{\psi }}(\mathbf{r}^{\prime },\mathbf{r},t)=\Phi _{0}^{*}(\mathbf{r}^{\prime })\Phi _{0}(\mathbf{r})+C_{\tilde{\psi }^{\dagger }\tilde{\psi }}(\mathbf{r}^{\prime },\mathbf{r},t)
\end{equation}
 and\begin{equation}
\label{symmetry-breaking-correlation-2}
C_{\hat{\psi }\hat{\psi }^{\dagger }}(\mathbf{r}^{\prime },\mathbf{r},t)=\Phi _{0}(\mathbf{r}^{\prime })\Phi _{0}^{*}(\mathbf{r})+C_{\tilde{\psi }\tilde{\psi }^{\dagger }}(\mathbf{r}^{\prime },\mathbf{r},t).
\end{equation}
 Using the detailed balance identity \cite{kadanoff-and-baym}\begin{equation}
\label{detailed-balance-identity}
C_{\tilde{\psi }\tilde{\psi }^{\dagger }}(\mathbf{r}^{\prime },\mathbf{r},\omega )=\mathrm{e}^{\beta \omega }C_{\tilde{\psi }^{\dagger }\tilde{\psi }}(\mathbf{r},\mathbf{r}^{\prime },-\omega ),
\end{equation}
 in addition to the Bose identity \begin{equation}
\label{Bose-function-identity}
\mathrm{e}^{\beta \omega }N^{0}(\omega )=1+N^{0}(\omega ),
\end{equation}
 one may easily write \( C_{\hat{\psi }^{\dagger }\hat{\psi }} \)
in terms of a spectral density (see p.\( \:  \)7 of Ref.\( \:  \)\cite{kadanoff-and-baym})
\begin{equation}
\label{correlation-relation-to-spectral-density}
C_{\tilde{\psi }^{\dagger }\tilde{\psi }}(\mathbf{r}^{\prime },\mathbf{r},-\omega )=N^{0}(\omega )A_{\tilde{\psi }^{\dagger }\tilde{\psi }}(\mathbf{r}^{\prime },\mathbf{r},\omega ).
\end{equation}
 Here \( N^{0}(\omega ) \) is the Bose distribution function \begin{equation}
\label{bose-distribution-function}
N^{0}(\omega )=\frac{1}{e^{\beta \omega }-1}
\end{equation}
 and the single-particle spectral density is \begin{equation}
\label{spectral-density-definition}
A_{\tilde{\psi }^{\dagger }\tilde{\psi }}(\mathbf{r}^{\prime },\mathbf{r},\omega )\equiv \int _{-\infty }^{\infty }\mathrm{d}t\mathrm{e}^{i\omega t}\left\langle \left[ \tilde{\psi }(\mathbf{r},t),\tilde{\psi }^{\dagger }(\mathbf{r}^{\prime })\right] \right\rangle _{eq}.
\end{equation}
 Note that for an inhomogeneous system, this spectral density is in
general not real.

In turn, \( A_{\tilde{\psi }^{\dagger }\tilde{\psi }} \) is given
by the difference of the retarded and advanced Green's functions,
\( G^{R} \) and \( G^{A} \) respectively, for the noncondensate
quantum field operators of a Bose gas (see p.\( \:  \)150 of Ref.\( \:  \)\cite{mahan}):
\begin{equation}
\label{spectral-density-relation-to-retarded-Green's-function}
A_{\tilde{\psi }^{\dagger }\tilde{\psi }}(\mathbf{r}^{\prime },\mathbf{r},\omega )=i\bigg (G^{R}(\mathbf{r},\mathbf{r}^{\prime },\omega +i\eta )-G^{A}(\mathbf{r},\mathbf{r}^{\prime },\omega -i\eta )\bigg ),
\end{equation}
 where \( G^{R}(\mathbf{r},\mathbf{r}^{\prime },\omega +i\eta ) \)
is the Fourier transform of \begin{equation}
\label{thermal retarded Green's function}
G^{R}(\mathbf{r},\mathbf{r}^{\prime },t)=-i\theta (t)\left\langle \left[ \tilde{\psi }(\mathbf{r},t),\tilde{\psi }^{\dagger }(\mathbf{r}^{\prime })\right] \right\rangle _{eq}
\end{equation}
 and \( G^{A}(\mathbf{r},\mathbf{r}^{\prime },\omega -i\eta ) \)
is the Fourier transform of \begin{equation}
\label{thermal advanced Green'sfunction}
G^{A}(\mathbf{r},\mathbf{r}^{\prime },t)=i\theta (-t)\left\langle \left[ \tilde{\psi }(\mathbf{r},t),\tilde{\psi }^{\dagger }(\mathbf{r}^{\prime })\right] \right\rangle _{eq}.
\end{equation}
The point of all this formalism is that standard finite-temperature
many-body theory \cite{fetter-walecka} most naturally determines
thermal Green's functions at imaginary Bose Matsubara frequencies
\( i\nu _{n}=i2n\pi /\beta ,\, n=0,\pm 1,\pm 2,\ldots  \). Fortunately,
one can determine \( G^{R,A}(\omega \pm i\eta ) \) by a unique analytic
continuation \( G(i\nu _{n}\rightarrow z\rightarrow \omega \pm i\eta )=G^{R,A}(\omega \pm i\eta ) \)
to the real frequency axis. This procedure thus gives the spectral
density \( A \).

We may express (\ref{deltaINa}) and (\ref{short time normal current})
in terms of the condensate order parameters and single-particle noncondensate
spectral densities. We find\begin{eqnarray}
\delta I_{Na} & = & \frac{\gamma ^{2}}{2\pi }\mathrm{Re}\int \mathrm{d}\mathbf{r}\int \mathrm{d}\mathbf{r}^{\prime }\mathrm{e}^{i\mathbf{q}\cdot (\mathbf{r}-\mathbf{r}^{\prime })}\nonumber \\
 &  & \Bigg \{i_{N}^{c-nc}(\mathbf{r},\mathbf{r}^{\prime },\omega -\Delta \varepsilon -\Delta \mu )\nonumber \\
 &  & +\int _{-\infty }^{\infty }\mathrm{d}\omega ^{\prime }i_{N}^{nc-nc}(\mathbf{r},\mathbf{r}^{\prime },\omega ^{\prime },\omega ^{\prime }+\omega -\Delta \varepsilon -\Delta \mu )\Bigg \}\label{deltaINa 1} 
\end{eqnarray}
 and\begin{eqnarray}
\delta I_{Nb} & = & -2\gamma ^{2}\, \mathrm{Im}\int \mathrm{d}\mathbf{r}\int \mathrm{d}\mathbf{r}^{\prime }\mathrm{e}^{i\mathbf{q}\cdot (\mathbf{r}-\mathbf{r}^{\prime })}\Bigg \{\frac{1}{2\pi }\mathrm{P}\int _{-\infty }^{\infty }\frac{\mathrm{d}\omega ^{\prime }}{2\pi }\frac{i_{N}^{c-nc}(\mathbf{r},\mathbf{r}^{\prime },\omega ^{\prime })}{\omega ^{\prime }-\omega +\Delta \varepsilon +\Delta \mu }\nonumber \\
 &  & +\mathrm{P}\int _{-\infty }^{\infty }\frac{\mathrm{d}\omega ^{\prime }}{2\pi }\int _{-\infty }^{\infty }\frac{\mathrm{d}\omega ^{\prime \prime }}{2\pi }\frac{i_{N}^{nc-nc}(\mathbf{r},\mathbf{r}^{\prime },\omega ^{\prime },\omega ^{\prime \prime })}{\omega ^{\prime \prime }-\omega ^{\prime }-\omega +\Delta \varepsilon +\Delta \mu }\Bigg \}.\label{deltaINb 1} 
\end{eqnarray}
 In these expressions we have introduced the normal current densities
\begin{equation}
\label{condensate-noncondensate current density}
i_{N}^{c-nc}(\mathbf{r},\mathbf{r}^{\prime },\omega )\equiv \Phi _{10}^{*}(\mathbf{r}^{\prime })\Phi _{10}(\mathbf{r})A_{\tilde{\psi }_{2}^{\dagger }\tilde{\psi }_{2}}(\mathbf{r},\mathbf{r}^{\prime },\omega )-\Phi _{20}^{*}(\mathbf{r})\Phi _{20}(\mathbf{r}^{\prime })A_{\tilde{\psi }_{1}^{\dagger }\tilde{\psi }_{1}}(\mathbf{r}^{\prime },\mathbf{r},-\omega ),
\end{equation}
 which corresponds to atomic tunneling of single particles between
the condensate and noncondensate components of the two gases. In contrast,
\begin{equation}
\label{noncondensate-noncondensate current density}
i_{N}^{nc-nc}(\mathbf{r},\mathbf{r}^{\prime },\omega ,\omega ^{\prime })\equiv \bigg (N_{1}^{0}(\omega )-N_{2}^{0}(\omega ^{\prime })\bigg )A_{\tilde{\psi }_{1}^{\dagger }\tilde{\psi }_{1}}(\mathbf{r}^{\prime },\mathbf{r},\omega )A_{\tilde{\psi }_{2}^{\dagger }\tilde{\psi }_{2}}(\mathbf{r},\mathbf{r}^{\prime },\omega ^{\prime }),
\end{equation}
 is associated with atomic tunneling of single atoms between the noncondensates
of the two gases.

We observe that in both \( \delta I_{Na} \) and \( \delta I_{Nb} \),
atoms make only two types of transitions between the states of the
two gases. The first transition process involves tunneling of an atom
between the condensate of one gas and the noncondensate of the other.
It is represented in \( \delta I_{Na} \) and \( \delta I_{Nb} \)
by terms of the form \( \Phi _{1(2)}^{*}\Phi _{1(2)}A_{\tilde{\psi }_{2(1)}^{\dagger }\tilde{\psi }_{2(1)}} \).
This contributes at all temperatures including \( T=0 \). The second
transition process involves the tunneling of an atom between the \emph{noncondensates}
of the two gases. It appears in \( \delta I_{Na} \) and \( \delta I_{Nb} \)
as a term of the form \( A_{\tilde{\psi }_{1}^{\dagger }\tilde{\psi }_{1}}A_{\tilde{\psi }_{2}^{\dagger }\tilde{\psi }_{2}} \).
This kind of transition can occur at zero temperature only when one
of the gases is Bose-condensed.

Finally we consider the {}``anomalous current'' given by the second
term in (\ref{linear response current 1}) in terms of the condensate
order parameters and noncondensate spectral density functions. We
follow the identical procedure used to find (\ref{deltaINa 1}) and
(\ref{deltaINb 1}) for the contributions \( \delta I_{Na} \) and
\( \delta I_{Nb} \) to the normal current. We obtain finally\begin{eqnarray}
\delta I_{A}(t) & \equiv  & 2\gamma ^{2}\, \mathrm{Re}\int \mathrm{d}\mathbf{r}\int \mathrm{d}\mathbf{r}^{\prime }\mathrm{e}^{i\mathbf{q}\cdot (\mathbf{r}+\mathbf{r}^{\prime })}\int _{-\infty }^{t}\mathrm{d}t^{\prime }\mathrm{e}^{-i(\omega -\Delta \varepsilon -\Delta \mu +i\eta )(t+t^{\prime })}\nonumber \\
 &  & \times \biggl [C_{\hat{\psi }_{1}\hat{\psi }_{1}}(\mathbf{r}^{\prime },\mathbf{r},t^{\prime }-t)C_{\hat{\psi }^{\dagger }_{2}\hat{\psi }^{\dagger }_{2}}(\mathbf{r}^{\prime },\mathbf{r},t^{\prime }-t)\nonumber \\
 &  & \qquad \qquad -C_{\hat{\psi }_{1}\hat{\psi }_{1}}(\mathbf{r},\mathbf{r}^{\prime },t-t^{\prime })C_{\hat{\psi }^{\dagger }_{2}\hat{\psi }^{\dagger }_{2}}(\mathbf{r},\mathbf{r}^{\prime },t-t^{\prime })\biggr ].\label{anomalous current definition} 
\end{eqnarray}
 Clearly this current may be written entirely in terms of the so-called
anomalous correlation functions involving creation or destruction
of \emph{pairs} of atoms, hence the name anomalous current. Using
\( t+t^{\prime }=2t+t^{\prime }-t\equiv 2t+\tau  \), we first rewrite
(\ref{anomalous current definition}) as \begin{eqnarray}
\delta I_{A}(t) & = & 2\gamma ^{2}\, \mathrm{Re}\Bigg \{\mathrm{e}^{-i2(\omega -\Delta \varepsilon -\Delta \mu )t}\int \mathrm{d}\mathbf{r}\int \mathrm{d}\mathbf{r}^{\prime }\mathrm{e}^{i\mathbf{q}\cdot (\mathbf{r}+\mathbf{r}^{\prime })}\int _{-\infty }^{\infty }\mathrm{d}\tau \theta (\tau )\mathrm{e}^{i(\omega -\Delta \varepsilon -\Delta \mu +i\eta )\tau }\nonumber \\
 &  & \times \biggl [C_{\hat{\psi }_{1}\hat{\psi }_{1}}(\mathbf{r}^{\prime },\mathbf{r},-\tau )C_{\hat{\psi }^{\dagger }_{2}\hat{\psi }^{\dagger }_{2}}(\mathbf{r}^{\prime },\mathbf{r},-\tau )-C_{\hat{\psi }_{1}\hat{\psi }_{1}}(\mathbf{r},\mathbf{r}^{\prime },\tau )C_{\hat{\psi }^{\dagger }_{2}\hat{\psi }^{\dagger }_{2}}(\mathbf{r},\mathbf{r}^{\prime },\tau )\biggr ]\Bigg \}.\label{anomalous current 1} 
\end{eqnarray}
 This shows that the anomalous current is explicitly \emph{time-dependent},
in contrast to the normal quasiparticle current given by (\ref{normal current 2}).
It clearly involves a second harmonic of the first-order Josephson
current in (\ref{equilibrium current 3}).

Following the same procedure we used in writing the normal current
in terms of the condensate order parameters and noncondensate spectral
density functions of the of the two gases, the anomalous current may
be rewritten as \begin{equation}
\label{decomposed anomalous current}
\delta I_{A}(t)\equiv \delta I_{Aa}(t)+\delta I_{Ab}(t),
\end{equation}
 where we have introduced the quantities \begin{eqnarray}
\delta I_{Aa}(t) & \equiv  & 2\gamma ^{2}\int \mathrm{d}\mathbf{r}\int \mathrm{d}\mathbf{r}^{\prime }J_{Aa}(\mathbf{r},\mathbf{r}^{\prime },\omega -\Delta \varepsilon -\Delta \mu )\nonumber \\
 &  & \qquad \qquad \qquad \times \cos \bigg [2(\omega -\Delta \varepsilon -\Delta \mu )t-\mathbf{q}\cdot (\mathbf{r}+\mathbf{r}^{\prime })\bigg ]\label{deltaIAa} 
\end{eqnarray}
 and \begin{eqnarray}
\delta I_{Ab}(t) & \equiv  & 2\gamma ^{2}\int \mathrm{d}\mathbf{r}\int \mathrm{d}\mathbf{r}^{\prime }J_{Ab}(\mathbf{r},\mathbf{r}^{\prime },\omega -\Delta \varepsilon -\Delta \mu )\nonumber \\
 &  & \qquad \qquad \qquad \times \sin \bigg [2(\omega -\Delta \varepsilon -\Delta \mu )t-\mathbf{q}\cdot (\mathbf{r}+\mathbf{r}^{\prime })\bigg ].\label{deltaIAb} 
\end{eqnarray}
 The quantities \( J_{Aa} \) and \( J_{Ab} \) in the above expressions
for \( \delta I_{Aa} \) and \( \delta I_{Ab} \) are given by \begin{eqnarray}
J_{Aa}(\mathbf{r},\mathbf{r}^{\prime },\omega -\Delta \varepsilon -\Delta \mu ) & \equiv  & \frac{1}{2}\int _{-\infty }^{\infty }\frac{\mathrm{d}\omega ^{\prime }}{2\pi }\mathrm{Re}\, i_{A}(\mathbf{r},\mathbf{r}^{\prime },\omega ^{\prime },\omega ^{\prime }-\omega +\Delta \varepsilon +\Delta \mu )\nonumber \\
 &  & -\mathrm{P}\int _{-\infty }^{\infty }\frac{\mathrm{d}\omega ^{\prime }}{2\pi }\int _{-\infty }^{\infty }\frac{\mathrm{d}\omega ^{\prime \prime }}{2\pi }\frac{\mathrm{Im}\, i_{A}(\mathbf{r},\mathbf{r}^{\prime },\omega ^{\prime },\omega ^{\prime \prime })}{\omega ^{\prime \prime }-\omega ^{\prime }+\omega -\Delta \varepsilon -\Delta \mu }\label{JA1} 
\end{eqnarray}
 and \begin{eqnarray}
J_{Ab}(\mathbf{r},\mathbf{r}^{\prime },\omega ) & \equiv  & \frac{1}{2}\int _{-\infty }^{\infty }\frac{\mathrm{d}\omega ^{\prime }}{2\pi }\mathrm{Im}\, i_{A}(\mathbf{r},\mathbf{r}^{\prime },\omega ^{\prime },\omega ^{\prime }-\omega +\Delta \varepsilon +\Delta \mu )\nonumber \\
 &  & +\mathrm{P}\int _{-\infty }^{\infty }\frac{\mathrm{d}\omega ^{\prime }}{2\pi }\int _{-\infty }^{\infty }\frac{\mathrm{d}\omega ^{\prime \prime }}{2\pi }\frac{\mathrm{Re}\, i_{A}(\mathbf{r},\mathbf{r}^{\prime },\omega ^{\prime },\omega ^{\prime \prime })}{\omega ^{\prime \prime }-\omega ^{\prime }+\omega -\Delta \varepsilon -\Delta \mu },\label{JA2} 
\end{eqnarray}
 where we have introduced the total anomalous current density \( i_{A} \),
which is written as \begin{eqnarray}
i_{A}(\mathbf{r},\mathbf{r}^{\prime },\omega ,\omega ^{\prime }) & \equiv  & \Phi _{10}(\mathbf{r})\Phi _{10}(\mathbf{r}^{\prime })A_{\tilde{\psi }_{2}^{\dagger }\tilde{\psi }_{2}^{\dagger }}(\mathbf{r},\mathbf{r}^{\prime },\omega ^{\prime })\delta (\omega )\nonumber \\
 &  & -\Phi _{20}^{*}(\mathbf{r})\Phi ^{*}_{20}(\mathbf{r}^{\prime })A_{\tilde{\psi }_{1}\tilde{\psi }_{1}}(\mathbf{r}^{\prime },\mathbf{r},\omega )\delta (\omega ^{\prime })\nonumber \\
 &  & +\bigg (N_{1}^{0}(\omega )-N_{2}^{0}(\omega ^{\prime })\bigg )A_{\tilde{\psi }_{1}\tilde{\psi }_{1}}(\mathbf{r}^{\prime },\mathbf{r},\omega )A_{\tilde{\psi }_{2}^{\dagger }\tilde{\psi }_{2}^{\dagger }}(\mathbf{r},\mathbf{r}^{\prime },\omega ^{\prime }).\label{total anomalous tunneling current} 
\end{eqnarray}
 Here \( A_{\tilde{\psi }\tilde{\psi }}(\mathbf{r}^{\prime },\mathbf{r},\omega ) \)
is the anomalous noncondensate single-particle spectral density function,
analogous to the noncondensate single-particle spectral density function
\( A_{\tilde{\psi }^{\dagger }\tilde{\psi }} \) defined in (\ref{spectral-density-definition}).

The first and second terms in (\ref{total anomalous tunneling current})
correspond to a tunneling current between the condensate of one gas
and the noncondensate of the other gas. The third term in (\ref{total anomalous tunneling current})
represents a tunneling current between the noncondensates of the two
gases.

One can rewrite \( J_{Aa} \) and \( J_{Ab} \) such that the contributions
from the condensate-noncondensate and noncondensate-noncondensate
tunneling processes are made more explicit. Thus, we have (see also
Ref.~\cite{finite-T-josephson}) \begin{eqnarray}
J_{Aa}(\mathbf{r},\mathbf{r}^{\prime },\omega -\Delta \varepsilon -\Delta \mu ) & = & \frac{1}{4\pi }\, \mathrm{Re}\, i_{A}^{c-nc}(\mathbf{r},\mathbf{r}^{\prime },\omega -\Delta \varepsilon -\Delta \mu )+\frac{1}{2\pi }\mathrm{P}\int _{-\infty }^{\infty }\frac{\mathrm{d}\omega ^{\prime }}{2\pi }\frac{\mathrm{Im}\, i_{A}^{c-nc}(\mathbf{r},\mathbf{r}^{\prime },\omega ^{\prime })}{\omega ^{\prime }-\omega +\Delta \varepsilon +\Delta \mu }\nonumber \\
 &  & +\frac{1}{2}\int _{-\infty }^{\infty }\frac{\mathrm{d}\omega ^{\prime }}{2\pi }\mathrm{Re}\, i^{nc-nc}_{A}(\mathbf{r},\mathbf{r}^{\prime },\omega ^{\prime },\omega ^{\prime }-\omega +\Delta \varepsilon +\Delta \mu )\nonumber \\
 &  & -\mathrm{P}\int _{-\infty }^{\infty }\frac{\mathrm{d}\omega ^{\prime }}{2\pi }\int _{-\infty }^{\infty }\frac{\mathrm{d}\omega ^{\prime \prime }}{2\pi }\frac{\mathrm{Im}\, i^{nc-nc}_{A}(\mathbf{r},\mathbf{r}^{\prime },\omega ^{\prime },\omega ^{\prime \prime })}{\omega ^{\prime \prime }-\omega ^{\prime }+\omega -\Delta \varepsilon -\Delta \mu }\label{JAa more explicit} 
\end{eqnarray}
 and \begin{eqnarray}
J_{Ab}(\mathbf{r},\mathbf{r}^{\prime },\omega -\Delta \varepsilon -\Delta \mu ) & = & \frac{1}{4\pi }\mathrm{Im}i_{A}^{c-nc}(\mathbf{r},\mathbf{r}^{\prime },\omega -\Delta \varepsilon -\Delta \mu )-\frac{1}{2\pi }\mathrm{P}\int _{-\infty }^{\infty }\frac{\mathrm{d}\omega ^{\prime }}{2\pi }\frac{\mathrm{Re}\, i_{A}^{c-nc}(\mathbf{r},\mathbf{r}^{\prime },\omega ^{\prime })}{\omega ^{\prime }-\omega +\Delta \varepsilon +\Delta \mu }\nonumber \\
 &  & +\frac{1}{2}\int _{-\infty }^{\infty }\frac{\mathrm{d}\omega ^{\prime }}{2\pi }\mathrm{Im}\, i^{nc-nc}_{A}(\mathbf{r},\mathbf{r}^{\prime },\omega ^{\prime },\omega ^{\prime }-\omega +\Delta \varepsilon +\Delta \mu )\nonumber \\
 &  & +\mathrm{P}\int _{-\infty }^{\infty }\frac{\mathrm{d}\omega ^{\prime }}{2\pi }\int _{-\infty }^{\infty }\frac{\mathrm{d}\omega ^{\prime \prime }}{2\pi }\frac{\mathrm{Re}\, i^{nc-nc}_{A}(\mathbf{r},\mathbf{r}^{\prime },\omega ^{\prime },\omega ^{\prime \prime })}{\omega ^{\prime \prime }-\omega ^{\prime }+\omega -\Delta \varepsilon -\Delta \mu },\label{JAb more explicit} 
\end{eqnarray}
 where we have introduced the condensate-noncondensate current density
(the analogue of (\ref{condensate-noncondensate current density}))
\begin{equation}
\label{condensate-noncondensate anomalous current density}
i_{A}^{c-nc}(\mathbf{r},\mathbf{r}^{\prime },\omega )\equiv \Phi _{10}(\mathbf{r})\Phi _{10}(\mathbf{r}^{\prime })A_{\tilde{\psi }_{2}^{\dagger }\tilde{\psi }_{2}^{\dagger }}(\mathbf{r},\mathbf{r}^{\prime },-\omega )-\Phi _{20}^{*}(\mathbf{r})\Phi ^{*}_{20}(\mathbf{r}^{\prime })A_{\tilde{\psi }_{1}\tilde{\psi }_{1}}(\mathbf{r}^{\prime },\mathbf{r},\omega ),
\end{equation}
 and the noncondensate-noncondensate current density (the analogue
of (\ref{noncondensate-noncondensate current density})) \begin{equation}
\label{noncondensate-noncondensate anomalous current density}
i_{A}^{nc-nc}(\mathbf{r},\mathbf{r}^{\prime },\omega ,\omega ^{\prime })\equiv \biggl (N_{1}^{0}(\omega )-N_{2}^{0}(\omega ^{\prime })\biggr )A_{\tilde{\psi }_{1}\tilde{\psi }_{1}}(\mathbf{r}^{\prime },\mathbf{r},\omega )A_{\tilde{\psi }_{2}^{\dagger }\tilde{\psi }_{2}^{\dagger }}(\mathbf{r},\mathbf{r}^{\prime },\omega ^{\prime }).
\end{equation}

We note that \( \delta I_{Ab} \) in (\ref{deltaIAb}) is the current
most directly associated with the \emph{coherent} tunneling responsible
for the Josephson effect in superconductors. This current is in-phase
with the Josephson current in (\ref{equilibrium current 3}) but involves
a second harmonic with a frequency \( 2(\omega -\Delta \varepsilon -\Delta \mu ) \).
It clarifies the role of the noncondensate in the BEC Josephson effect.
On the other hand, the \( \delta I_{Aa} \) current in (\ref{deltaIAa})
is directly related to a \emph{dissipative} interference current,
just as one finds for the Josephson effect in superconductors \cite{barone,finite-T-josephson}.

\section{Coupling between two trapped gases}

\label{Section: Coupling between two trapped gases}In this section,
we use the results of Section \ref{Section: Tunneling currents in terms of correlations functions}
to calculate the second order tunneling current involved in a Raman
scattering experiment. The spectral density functions are calculated
in the finite temperature Bogoliubov-Popov approximation \cite{griffin-gapless}.
The Josephson tunneling current has already been discussed and is
given by (\ref{equilibrium current 3}).

\begin{figure*}[p]
{\centering \resizebox*{0.6\textwidth}{0.4\textheight}{\includegraphics{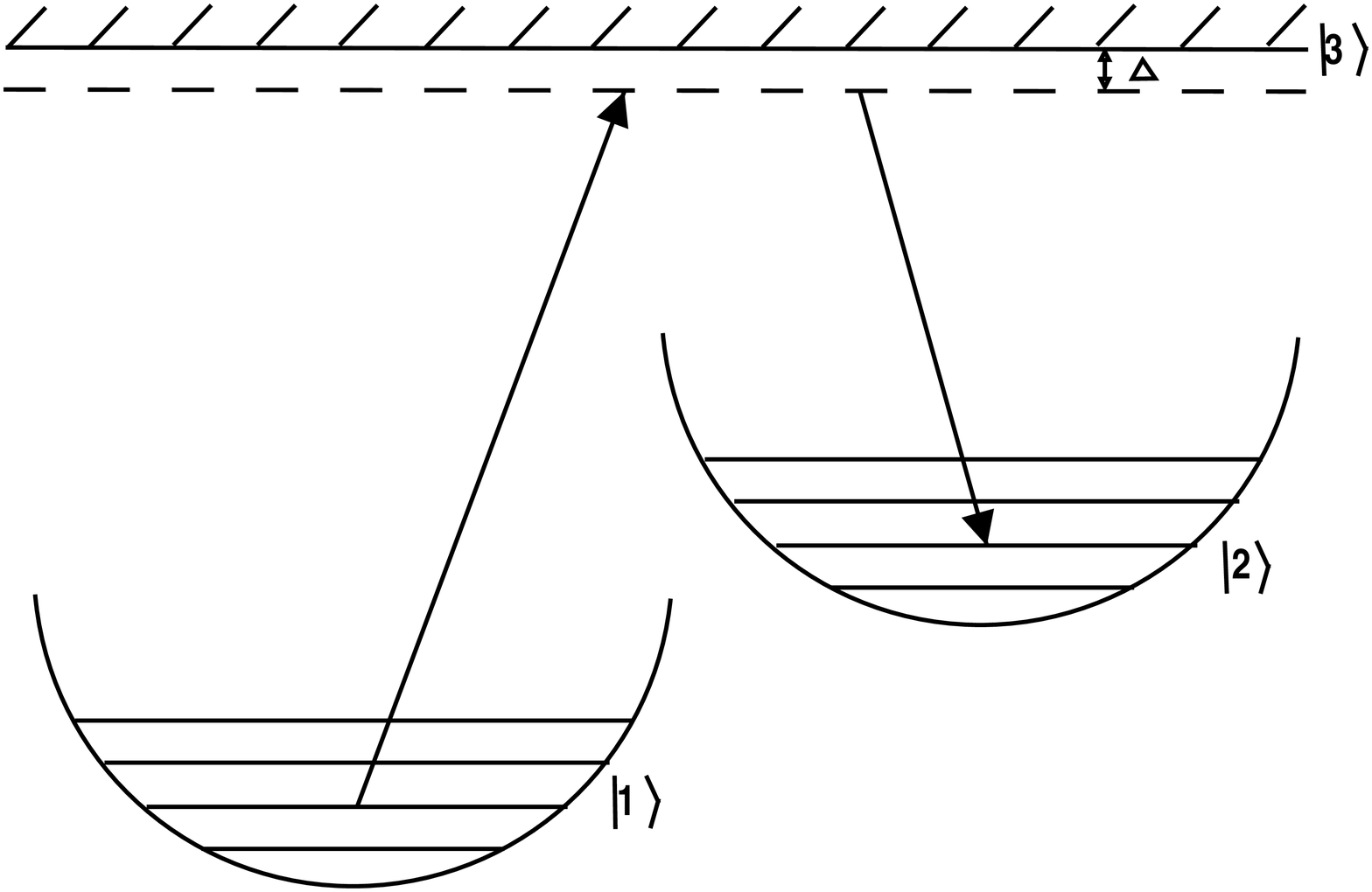}} \par}

\caption{\label{fig:raman-figure}Raman process coupling two atomic hyperfine
levels that may or may not be Bose-condensed.}
\end{figure*}

A two-photon Bragg scattering experiment involves two laser beams
at different frequencies and with different propagation vectors \cite{structure-factor-phonons-houches-exp}.
However, we are interested in a different type of two-photon process
in which an atom changes its hyperfine state. This is a Raman process
and is depicted in Fig.\( \:  \)\ref{fig:raman-figure}. We now discuss
how the effective frequency \( \omega  \) and wavevector \( \mathbf{q} \)
in (\ref{V}) are defined in such a Raman scattering experiment \cite{raman-output-coupler}.
Assuming that each laser beam has equal intensity \( I \), the combined
intensity of the two beams at space-time point \( (\mathbf{r},t) \)
is given by \begin{equation}
\label{modulated-bragg-intensity}
I_{tot}(\mathbf{r},t)=2I+2I\cos (\mathbf{q}\cdot \mathbf{r}-\omega t),
\end{equation}
 where \( \omega  \) is the frequency difference between the two
laser beams\begin{equation}
\label{bragg-Delta}
\omega \equiv \omega _{a}-\omega _{b}.
\end{equation}
 The wavevector difference between the two laser beams is \begin{equation}
\label{bragg q vector}
\mathbf{q}=\mathbf{k}_{a}-\mathbf{k}_{b},
\end{equation}
 where \( \mathbf{k}_{j} \) is the wavevector of laser beam \( j \).
We may take \( k_{1}^{2}=k_{2}^{2}=\bar{k}^{2} \) such that the standard
Bragg scattering relation (see, for example, p.\( \:  \)103 of Ref.\( \:  \)\cite{ashcroft-mermin})
holds and \( q \) is given by \begin{equation}
\label{bragg-conservation-relation}
q^{2}=4\bar{k}^{2}\sin ^{2}\left( \frac{\theta }{2}\right) ,
\end{equation}
 where \( \theta  \) is the angle between the two counter-propagating
beams. Finally the coupling parameter \( \gamma  \) in (\ref{V})
is related to the single-photon Rabi frequency \( \Omega _{Rabi} \)
(assumed to be the same for both hyperfine level transitions) and
the laser beam detuning \( \Delta  \) from the intermediate level
\( \left| 3\right\rangle  \) (see Fig.~\ref{fig:raman-figure})
according to \cite{raman-output-coupler} \begin{equation}
\label{coupling parameter for Braggscattering experiment}
\gamma \propto \frac{\left| \Omega _{Rabi}\right| ^{2}}{\Delta }\propto \Omega _{Rabi}^{(2)},
\end{equation}
 where \( \Omega _{Rabi}^{(2)} \) is the so-called two-photon {}``Rabi
frequency\char`\"{}.

The many-body Hamiltonian (in the absence of a tunneling perturbation)
is \( \hat{H}_{0}=\hat{H}_{10}+\hat{H}_{20} \), where \begin{equation}
\label{H10}
\hat{H}_{10}=\int \mathrm{d}\mathbf{r}\left\{ \hat{\psi }_{1}^{\dagger }(\mathbf{r})\left[ -\frac{\nabla ^{2}}{2m}+U_{ext}(\mathbf{r})\right] \hat{\psi }_{1}(\mathbf{r})+\frac{g_{1}}{2}\hat{\psi }_{1}^{\dagger }(\mathbf{r})\hat{\psi }_{1}^{\dagger }(\mathbf{r})\hat{\psi }_{1}(\mathbf{r})\hat{\psi }_{1}(\mathbf{r})\right\} .
\end{equation}
 In this expression for \( \hat{H}_{10} \), \( U_{ext} \) is the
trapping potential and \( g_{1}=4\pi a_{1}/m \) is the s-wave scattering
pseudo-potential with \( a_{1} \) being the s-wave scattering length
for atoms in gas \( 1 \). The expression for \( \hat{H}_{20} \)
has an analogous form. We recall that the spectral densities in Section
\ref{Section: Tunneling currents in terms of correlations functions}
are defined in terms of the grand canonical Hamiltonian \( \hat{K}_{i0}\equiv \hat{H}_{i0}-\mu _{i}\hat{N}_{i} \).
Thus the excitation energies of gas \( i \) are measured relative
to the chemical potential \( \mu _{i} \).

\subsection{\label{sec:The Finite T Bogoliubov Approximation}The Bogoliubov-Popov
approximation}

In the standard static Bogoliubov-Popov approximation for excitations
in a trapped Bose gas \cite{fetter-walecka,griffin-gapless,rmp-bose-gas-review},
the noncondensate field operator \( \tilde{\psi } \) can be written
as\begin{equation}
\label{bogoliubov-quasi-particle-quantization}
\tilde{\psi }(\mathbf{r},t)=\sum _{j}\left( u_{j}(\mathbf{r})\mathrm{e}^{-iE_{j}t}\hat{\alpha }_{j}-v^{*}_{j}(\mathbf{r})\mathrm{e}^{iE_{j}t}\hat{\alpha }^{\dagger }_{j}\right) ,
\end{equation}
 where \( u_{j}(\mathbf{r}) \) and \( v_{j}(\mathbf{r}) \) are the
Bogoliubov quasiparticle amplitudes, \( E_{j} \) is the energy of
Bogoliubov state \( j \) and \( \hat{\alpha }_{j} \) (\( \hat{\alpha }^{\dagger }_{j} \))
annihilate (create) a quasiparticle in state \( j \), respectively.
These annihilation and creation operators are defined to satisfy the
boson commutation relations\begin{equation}
\label{bogoliubov-commutation-relations}
\left[ \hat{\alpha }_{m},\hat{\alpha }_{n}\right] =\left[ \hat{\alpha }_{m}^{\dagger },\hat{\alpha }_{n}^{\dagger }\right] =0;\qquad \left[ \hat{\alpha }_{m},\hat{\alpha }_{n}^{\dagger }\right] =\delta _{mn}.
\end{equation}
 The Bogoliubov quasiparticle amplitudes must satisfy the finite temperature
coupled Bogoliubov-Popov equations of motion \cite{griffin-gapless}\begin{eqnarray}
\hat{\mathcal{L}}u_{j}(\mathbf{r})-gn_{c}(\mathbf{r})v_{j}(\mathbf{r}) & = & E_{j}u_{j}(\mathbf{r})\nonumber \label{u-equation-of-motion} \\
\hat{\mathcal{L}}v_{j}(\mathbf{r})-gn_{c}(\mathbf{r})u_{j}(\mathbf{r}) & = & -E_{j}v_{j}(\mathbf{r})\label{bogoliubov-equations-of-motion} 
\end{eqnarray}
 in order for \( \hat{K}_{0} \) to be reduced to the diagonal form
\begin{equation}
\label{diagonalized Hamiltonian in Bogoliubov-Popov approximation}
\hat{K}^{Bog}=const+\sum _{j}E_{j}\hat{\alpha }_{j}^{\dagger }\hat{\alpha }_{j}.
\end{equation}
 In (\ref{bogoliubov-equations-of-motion}), we have introduced the
differential operator\begin{equation}
\label{bogoliubov-differential-operator}
\hat{\mathcal{L}}\equiv -\frac{\nabla ^{2}}{2m}+U_{ext}(\mathbf{r})-\mu +2gn(\mathbf{r}),
\end{equation}
 where \( n(\mathbf{r}) \) is the total static density.

Using the commutation relations (\ref{bogoliubov-commutation-relations}),
one finds that the single-particle spectral density function can be
expressed in terms of Bogoliubov quasiparticles as\begin{equation}
\label{bogoliubov-spectral-density-time}
A_{\tilde{\psi }^{\dagger }\tilde{\psi }}(\mathbf{r}^{\prime },\mathbf{r},t-t^{\prime })=\sum _{j}\Biggl [u^{*}_{j}(\mathbf{r}^{\prime })u_{j}(\mathbf{r})\mathrm{e}^{-iE_{j}(t-t^{\prime })}-v_{j}(\mathbf{r}^{\prime })v^{*}_{j}(\mathbf{r})\mathrm{e}^{iE_{j}(t-t^{\prime })}\Biggr ],
\end{equation}
 which has the Fourier transform\begin{equation}
\label{bogoliubov-spectral-density-frequency}
A_{\tilde{\psi }^{\dagger }\tilde{\psi }}(\mathbf{r}^{\prime },\mathbf{r},\omega )=2\pi \sum _{j}\Biggl [u^{*}_{j}(\mathbf{r}^{\prime })u_{j}(\mathbf{r})\delta (\omega -E_{j})-v_{j}(\mathbf{r}^{\prime })v^{*}_{j}(\mathbf{r})\delta (\omega +E_{j})\Biggr ].
\end{equation}
 We note the appearance in (\ref{bogoliubov-spectral-density-frequency})
of negative energy poles (\( \omega =-E_{j} \)) in addition to the
usual positive energy poles (\( \omega =E_{j} \)). These are a result
of unique processes induced by the Bose condensate. The two poles
follow immediately from (\ref{bogoliubov-quasi-particle-quantization}),
which shows how the destruction of an atom involves a linear superposition
of the destruction (amplitude \( u \)) and creation (amplitude \( v \))
of Bogoliubov excitations.

Similarly, the off-diagonal or anomalous single-particle spectral
density function of relevance to the anomalous current densities \( i^{c-nc}_{A} \)
in (\ref{condensate-noncondensate anomalous current density}) and
\( i_{A}^{nc-nc} \) in (\ref{noncondensate-noncondensate anomalous current density})
is given by \
\begin{equation}
\label{bogoliubov spectral density for anomalous correlations}
A_{\tilde{\psi }\tilde{\psi }}(\mathbf{r}^{\prime },\mathbf{r},\omega )=-2\pi \sum _{j}\Biggl [u_{j}(\mathbf{r})v_{j}^{*}(\mathbf{r}^{\prime })\delta (\omega -E_{j})-u_{j}(\mathbf{r}^{\prime })v_{j}^{*}(\mathbf{r})\delta (\omega +E_{j})\Biggr ].
\end{equation}
 The spectral density functions can be shown to satisfy \( A_{\tilde{\psi }^{\dagger }\tilde{\psi }^{\dagger }}(\mathbf{r}^{\prime },\mathbf{r},\omega )=A^{*}_{\tilde{\psi }\tilde{\psi }}(\mathbf{r},\mathbf{r}^{\prime },\omega ) \)
and \( A_{\tilde{\psi }^{\dagger }\tilde{\psi }}(\mathbf{r}^{\prime },\mathbf{r},\omega )=A^{*}_{\tilde{\psi }\tilde{\psi }^{\dagger }}(\mathbf{r},\mathbf{r}^{\prime },\omega ) \).

\subsection{\label{sec:The Normal and Anomalous Current in the Bogoliubov Approximation}The
normal and anomalous currents in the Bogoliubov-Popov approximation}

Using the results in (\ref{bogoliubov-spectral-density-frequency})
and (\ref{bogoliubov spectral density for anomalous correlations}),
we can now write the normal and anomalous linear response currents
in terms of Bogoliubov excitations. In Section \ref{Section: Tunneling currents in terms of correlations functions},
we found that two distinct physical processes contribute to the normal
current. The current density of atoms tunneling between the condensates
and noncondensates of the two gases is given by (\ref{condensate-noncondensate current density}),
or \begin{eqnarray}
i_{N}^{c-nc}(\mathbf{r},\mathbf{r}^{\prime },\omega ) & = & 2\pi \sum _{j}\Bigg [\Phi ^{*}_{10}(\mathbf{r})\Phi _{10}(\mathbf{r}^{\prime })u_{2j}^{*}(\mathbf{r})u_{2j}(\mathbf{r}^{\prime })\delta (\omega -E_{2j})\nonumber \\
 &  & -\Phi _{10}^{*}(\mathbf{r}^{\prime })\Phi _{10}(\mathbf{r})v_{2j}(\mathbf{r}^{\prime })v^{*}_{2j}(\mathbf{r})\delta (\omega +E_{2j})\nonumber \\
 &  & -\Phi ^{*}_{20}(\mathbf{r}^{\prime })\Phi _{20}(\mathbf{r})u_{1j}^{*}(\mathbf{r}^{\prime })u_{1j}(\mathbf{r})\delta (\omega +E_{1j})\nonumber \\
 &  & +\Phi _{20}^{*}(\mathbf{r})\Phi _{20}(\mathbf{r}^{\prime })v_{1j}(\mathbf{r})v^{*}_{1j}(\mathbf{r}^{\prime })\delta (\omega -E_{1j})\Bigg ].\label{condensate-noncondensate normal current density in Bogoliubov approximation} 
\end{eqnarray}
 This type of process involves both condensate atoms and quasiparticles.
The current density associated with the tunneling of atoms between
the noncondensates of the two gases has a more complicated form given
by (\ref{noncondensate-noncondensate current density}), or \begin{eqnarray}
i_{N}^{nc-nc}(\mathbf{r},\mathbf{r}^{\prime },\omega ,\omega ^{\prime }) & = & (2\pi )^{2}\biggl (N_{1}^{0}(\omega )-N_{2}^{0}(\omega ^{\prime })\biggr )\nonumber \\
 &  & \sum _{i,j}\Biggl [u^{*}_{1i}(\mathbf{r}^{\prime })u_{1i}(\mathbf{r})u_{2j}^{*}(\mathbf{r})u_{2j}(\mathbf{r}^{\prime })\delta (\omega -E_{1i})\delta (\omega ^{\prime }-E_{2j})\nonumber \\
 &  & -u_{1i}^{*}(\mathbf{r}^{\prime })u_{1i}(\mathbf{r})v_{2j}(\mathbf{r})v_{2j}^{*}(\mathbf{r}^{\prime })\delta (\omega -E_{1i})\delta (\omega ^{\prime }+E_{2j})\nonumber \\
 &  & -v_{1i}(\mathbf{r}^{\prime })v_{1i}^{*}(\mathbf{r})u_{2j}^{*}(\mathbf{r})u_{2j}(\mathbf{r}^{\prime })\delta (\omega +E_{1i})\delta (\omega ^{\prime }-E_{2j})\nonumber \\
 &  & +v_{1i}(\mathbf{r}^{\prime })v_{1i}^{*}(\mathbf{r})v_{2j}(\mathbf{r})v^{*}_{2j}(\mathbf{r}^{\prime })\delta (\omega +E_{1i})\delta (\omega ^{\prime }+E_{2j})\Biggr ].\label{noncondensate-noncondensate normal current density in Bogoliubov approximation} 
\end{eqnarray}

As in the case of the normal current, two different physical processes
contribute to the anomalous current in (\ref{anomalous current definition}).
The first process involves pairs of condensate atoms tunneling to
noncondensate states of the other Bose gas. This anomalous current
density is given by (\ref{condensate-noncondensate anomalous current density}),
namely \begin{eqnarray}
i_{A}^{c-nc}(\mathbf{r},\mathbf{r}^{\prime },\omega ) & = & 2\pi \sum _{j}\Biggl [\Phi _{10}(\mathbf{r}^{\prime })\Phi _{10}(\mathbf{r})u_{2j}(\mathbf{r})v^{*}_{2j}(\mathbf{r}^{\prime })\delta (\omega +E_{2j})\nonumber \\
 &  & -\Phi ^{*}_{10}(\mathbf{r}^{\prime })\Phi ^{*}_{10}(\mathbf{r})u^{*}_{2j}(\mathbf{r}^{\prime })v_{2j}(\mathbf{r})\delta (\omega -E_{2j})\nonumber \\
 &  & -\Phi ^{*}_{20}(\mathbf{r})\Phi ^{*}_{20}(\mathbf{r}^{\prime })u_{1j}(\mathbf{r}^{\prime })v^{*}_{1j}(\mathbf{r})\delta (\omega +E_{1j})\nonumber \\
 &  & +\Phi ^{*}_{20}(\mathbf{r})\Phi ^{*}_{20}(\mathbf{r}^{\prime })u^{*}_{1j}(\mathbf{r})v_{1j}(\mathbf{r}^{\prime })\delta (\omega -E_{1j})\Biggr ].\label{condensate-noncondensate anomalous current density in Bogoliubov approximation} 
\end{eqnarray}
 The expression for the anomalous current density between condensate
1 and noncondensate 2 has a similar form. The final process involves
quasiparticles of both gases and is given by (\ref{noncondensate-noncondensate anomalous current density}),
namely\begin{eqnarray}
i_{A}^{nc-nc}(\mathbf{r},\mathbf{r}^{\prime },\omega ,\omega ^{\prime }) & = & (2\pi )^{2}\biggl (N_{1}^{0}(\omega )-N_{2}^{0}(\omega ^{\prime })\biggr )\nonumber \\
 &  & \sum _{i,j}\Biggl [u_{1i}(\mathbf{r}^{\prime })v_{1i}^{*}(\mathbf{r})u_{2j}^{*}(\mathbf{r}^{\prime })v_{2j}(\mathbf{r})\delta (\omega +E_{1i})\delta (\omega ^{\prime }+E_{2j})\nonumber \\
 &  & -u_{1i}(\mathbf{r}^{\prime })v_{1i}^{*}(\mathbf{r})u_{2j}(\mathbf{r})v_{2j}^{*}(\mathbf{r}^{\prime })\delta (\omega +E_{1i})\delta (\omega ^{\prime }-E_{2j})\nonumber \\
 &  & -u^{*}_{1i}(\mathbf{r})v_{1i}(\mathbf{r}^{\prime })u_{2j}^{*}(\mathbf{r}^{\prime })v_{2j}(\mathbf{r})\delta (\omega -E_{1i})\delta (\omega ^{\prime }+E_{2j})\nonumber \\
 &  & +u^{*}_{1i}(\mathbf{r})v_{1i}(\mathbf{r}^{\prime })u_{2j}(\mathbf{r})v_{2j}^{*}(\mathbf{r}^{\prime })\delta (\omega -E_{1i})\delta (\omega ^{\prime }-E_{2j})\Biggr ].\label{noncondensate-noncondensate anomalous current density in Bogoliubov approximation} 
\end{eqnarray}

As with the Josephson current given in (\ref{equilibrium current 3}),
one sees from (\ref{condensate-noncondensate normal current density in Bogoliubov approximation})-(\ref{noncondensate-noncondensate anomalous current density in Bogoliubov approximation})
that the magnitude of the normal and anomalous quasiparticle currents
is determined by the spatial overlap of the Bogoliubov wavefunctions
described by \( u_{i}(\mathbf{r}) \) and \( v_{i}(\mathbf{r}) \)
for the two systems.

When one uses (\ref{condensate-noncondensate normal current density in Bogoliubov approximation})-(\ref{noncondensate-noncondensate anomalous current density in Bogoliubov approximation})
in the expressions for the normal (see (\ref{deltaINa 1}) and (\ref{deltaINb 1}))
and anomalous (see (\ref{deltaIAa}) and (\ref{deltaIAb})) currents,
our results are essentially equivalent to expressions discussed in
Ref.~\cite{finite-T-josephson}. The main differences are that Meier
and Zwerger approximated the two Bose gases as being uniform (or homogeneous)
and used the single-particle spectral densities at \( T=0 \).

\section{Out-coupling to a non-Bose-condensed gas}

\label{sec:Case of Out-coupling to a non-Bose-condensed Gas}In this
section, we consider the special case of atoms out-coupling to the
vacuum as discussed in Refs.~\cite{burnett-1,burnett-2}. More precisely,
we first deal with the case when gas 2 is a uniform non-interacting
gas. The corresponding spectral density function then has the well-known
form \cite{kadanoff-and-baym,fetter-walecka}\begin{equation}
\label{uniform-gas-spectral-density}
A_{\hat{\psi }_{2}^{\dagger }\hat{\psi }_{2}}(\mathbf{k},\omega )=2\pi \delta \left( \omega -\tilde{\varepsilon }_{\mathbf{k}2}\right) ,
\end{equation}
 where\begin{equation}
\label{gas2-atom-excitation-energy}
\tilde{\varepsilon }_{\mathbf{k}2}\equiv k^{2}/2m-\mu _{2}
\end{equation}
 is the the excitation energy of an atom in gas \( 2 \). Because
the gas is assumed to be in the classical limit, we have \( N_{2}^{0}(\tilde{\varepsilon }_{\mathbf{k}2})\ll 1 \).

The only terms in (\ref{condensate-noncondensate normal current density in Bogoliubov approximation})
which are finite reduce to \begin{eqnarray}
i_{Nout}^{c-nc}(\mathbf{r},\mathbf{r}^{\prime },\omega ) & = & \Phi _{10}^{*}(\mathbf{r}^{\prime })\Phi _{10}(\mathbf{r})A_{\tilde{\psi }_{2}^{\dagger }\tilde{\psi }_{2}}(\mathbf{r}-\mathbf{r}^{\prime },\omega )\nonumber \\
 & = & 2\pi \int \frac{\mathrm{d}\mathbf{k}}{(2\pi )^{3}}\mathrm{e}^{-i\mathbf{k}\cdot (\mathbf{r}-\mathbf{r}^{\prime })}\Phi _{10}^{*}(\mathbf{r}^{\prime })\Phi _{10}(\mathbf{r})\delta (\tilde{\varepsilon }_{\mathbf{k}2}-\omega )\label{outcoupling current density condensate-noncondensate} 
\end{eqnarray}
 and in (\ref{noncondensate-noncondensate normal current density in Bogoliubov approximation})\begin{eqnarray}
i_{Nout}^{nc-nc}(\mathbf{r},\mathbf{r}^{\prime },\omega ,\omega ^{\prime }) & = & N_{1}^{0}(\omega )A_{\tilde{\psi }_{1}^{\dagger }\tilde{\psi }_{1}}(\mathbf{r}^{\prime },\mathbf{r},\omega )A_{\tilde{\psi }_{2}^{\dagger }\tilde{\psi }_{2}}(\mathbf{r}-\mathbf{r}^{\prime },\omega ^{\prime })\nonumber \\
 & = & 4\pi ^{2}\sum _{j}\int \frac{\mathrm{d}\mathbf{k}}{(2\pi )^{3}}\biggl \{N_{1}^{0}(E_{1j})u^{*}_{1j}(\mathbf{r}^{\prime })u_{1j}(\mathbf{r})\delta \left( \omega -E_{1j}\right) \nonumber \\
 &  & +[1+N_{1}^{0}(E_{1j})]v_{1j}(\mathbf{r}^{\prime })v^{*}_{1j}(\mathbf{r})\delta \left( \omega +E_{1j}\right) \biggr \}\delta (\tilde{\varepsilon }_{\mathbf{k}2}-\omega ^{\prime }).\label{outcoupling current density noncondensate-noncondensate} 
\end{eqnarray}
 We have used the Bose identity \( N^{0}(-\omega )=-[1+N^{0}(\omega )] \)
in writing the last term in (\ref{outcoupling current density noncondensate-noncondensate}).
The condition that \( N_{2}^{0}\ll 1 \) results in negligible tunneling
from gas 2 back to gas 1.

The current density \( i_{Nout}^{(c-nc)} \) in (\ref{outcoupling current density condensate-noncondensate})
corresponds to the \emph{destruction} of a condensate atom in gas
\( 1 \) and the \emph{creation} of an excitation in gas \( 2 \)
with energy \( \tilde{\varepsilon }_{\mathbf{k}2} \). The first term
in the density \( i_{Nout}^{(nc-nc)} \) of (\ref{outcoupling current density noncondensate-noncondensate})
represents what might be called the {}``quantum evaporation'' of
an atom, \emph{destroying} a Bogoliubov quasiparticle with energy
\( E_{1j}=k^{2}/2m-(\omega -\Delta \varepsilon +\mu _{1}) \) and
\emph{creating} an atom in gas \( 2 \) with \( k^{2}/2m \). The
second term in (\ref{outcoupling current density noncondensate-noncondensate})
corresponds to \emph{creating} an atom in gas \( 2 \) with energy
\( k^{2}/2m \) and simultaneously \emph{creating} a quasiparticle
excitation in gas \( 1 \) with energy \( E_{1j}=-[k^{2}/2m-(\omega -\Delta \varepsilon +\mu _{1})] \).
This second out-coupling channel in (\ref{outcoupling current density noncondensate-noncondensate})
is a direct consequence of the unusual condensate-induced correlations
in a Bose gas. It corresponds to the {}``pair-breaking'' process
discussed by the Oxford group \cite{burnett-1}. The analogue of this
process appears in the predicted lineshape of the recombination photons
emitted in the decay of an exciton in a Bose-condensed gas of optically-excited
excitons \cite{svg}. All three out-coupling contributions will be
calculated in Section \ref{sec:LDA}.

Using (\ref{outcoupling current density condensate-noncondensate})
and (\ref{outcoupling current density noncondensate-noncondensate}),
the out-coupling current \( \delta I_{out} \) is\begin{eqnarray}
\delta I_{out} & = & \gamma ^{2}\int \mathrm{d}\mathbf{r}\int \mathrm{d}\mathbf{r}^{\prime }\int \frac{\mathrm{d}\mathbf{k}}{(2\pi )^{3}}\mathrm{e}^{i(\mathbf{q}-\mathbf{k})\cdot (\mathbf{r}-\mathbf{r}^{\prime })}\nonumber \\
 &  & \qquad \Biggl \{\Phi _{10}^{*}(\mathbf{r}^{\prime })\Phi _{10}(\mathbf{r})\delta \left( k^{2}/2m-(\omega -\Delta \varepsilon +\mu _{1})\right) \nonumber \\
 &  & +2\pi \sum _{j}\biggl [N_{1}^{0}(E_{1j})u_{1j}^{*}(\mathbf{r}^{\prime })u_{1j}(\mathbf{r})\delta \left( k^{2}/2m-(\omega -\Delta \varepsilon +\mu _{1})-E_{1j}\right) \nonumber \\
 &  & \qquad +[1+N_{1}^{0}(E_{1j})]v_{1j}(\mathbf{r}^{\prime })v^{*}_{1j}(\mathbf{r})\delta \left( k^{2}/2m-(\omega -\Delta \varepsilon +\mu _{1})+E_{1j}\right) \biggr ]\Biggr \}.\label{bogoliubov outcoupling current} 
\end{eqnarray}
 We refer to the second term as the {}``\( u^{2} \) contribution''
and the third term as the {}``\( v^{2} \) contribution''. It is
useful to describe the main features of the out-coupling current spectrum
arising from the three terms in (\ref{bogoliubov outcoupling current}),
as a function of the energy \( k^{2}/2m \) of the out-coupled atoms,
for fixed \( \omega  \) and \textbf{\( \mathbf{q} \)}. The energy
conserving delta functions in (\ref{bogoliubov outcoupling current})
determine the energies of the out-coupling current peaks for the different
processes. There is a delta function peak at \( k^{2}/2m+\varepsilon _{2}=\omega +\varepsilon _{1}+\mu _{1} \)
from atoms coming directly from the condensate, where \( \mu _{1} \)
is the condensate energy of a trapped atom in the hyperfine state
\( \left| 1\right\rangle  \) and \( \varepsilon _{i} \) is the hyperfine
state energy. Relative to this condensate pole, the peak associated
with the quantum evaporation term is shifted by a positive amount
\( E_{1j} \), while the peak of the {}``pair-breaking'' channel
is shifted by a negative amount \( -E_{1j} \). These shifts follow
from the fact that the quantum evaporation channel arises from the
positive energy pole (related to \( u^{2} \)) and the pair-breaking
channel corresponds to the negative energy pole (related to \( v^{2} \))
in the single-particle spectral density.

We next consider the behavior of the three out-coupling current contributions
in (\ref{bogoliubov outcoupling current}) as a function of temperature.
For \( T_{1}=0 \), since \( N_{1}^{0}(E_{1j})\rightarrow 0 \), (i.e.~no
thermally excited quasiparticles) the second channel (\( u^{2} \))
has zero weight. Thus, at \( T_{1}=0 \), only the first and third
processes occur. When gas 1 is at a finite temperature (\( T_{1}\neq 0 \)),
the second channel (\( u^{2} \)) begins to contribute to the out-coupling
current, taking weight away from the pair-breaking channel (\( v^{2} \)).
All these features are shown explicitly in the numerical results given
in Section \ref{sec:LDA} (see Figs.~\ref{fig:condensate-noncondensate out-coupling current for q=3D3D2/aho and a=3D3D5nm}-\ref{fig:total out-coupling current for q=3D3D10/aho a=3D3D5nm T=3D3D0.2TBEC}).

The out-coupling current derived in Ref.~\cite{burnett-2} is given
as the rate of atoms excited out of gas 1 with energy \( \tilde{\varepsilon }_{\mathbf{k}2} \)
and may be written as (where we use a notation more consistent with
our own)\begin{eqnarray}
\frac{\mathrm{d}n_{\mathbf{k}}}{\mathrm{d}t} & = & \gamma ^{2}\int \mathrm{d}\mathbf{r}\int \mathrm{d}\mathbf{r}^{\prime }\mathrm{e}^{i\mathbf{q}\cdot (\mathbf{r}-\mathbf{r}^{\prime })}\varphi ^{*}_{\mathbf{k}2}(\mathbf{r})\varphi _{\mathbf{k}2}(\mathbf{r}^{\prime })\nonumber \\
 &  & \qquad \times \Biggl \{\Phi ^{*}_{10}(\mathbf{r}^{\prime })\Phi _{10}(\mathbf{r})\delta \left( k^{2}/2m-(\omega -\Delta \varepsilon +\mu _{1})\right) \nonumber \\
 &  & +2\pi \sum _{j}\biggl [N^{0}_{1}(E_{1j})u^{*}_{1j}(\mathbf{r}^{\prime })u_{1j}(\mathbf{r})\delta \left( k^{2}/2m-(\omega -\Delta \varepsilon +\mu _{1})-E_{1j}\right) \nonumber \\
 &  & \qquad +[1+N^{0}_{1}(E_{1j})]v_{1j}(\mathbf{r}^{\prime })v^{*}_{1j}(\mathbf{r})\delta \left( k^{2}/2m-(\omega -\Delta \varepsilon +\mu _{1})+E_{1j}\right) \biggr ]\Biggr \},\label{oxford-result} 
\end{eqnarray}
 where \( \varphi _{\mathbf{k}2}(\mathbf{r}) \) is the wavefunction
for the out-coupled excitation with energy \( \tilde{\varepsilon }_{\mathbf{k}2} \).
We can make a direct comparison between the integrands of (\ref{bogoliubov outcoupling current})
and (\ref{oxford-result}). Indeed taking \( \varphi _{2\mathbf{k}}(\mathbf{r})=e^{i\mathbf{k}\cdot \mathbf{r}} \)
(a free atom with momentum \( \mathbf{k} \)) reduces (\ref{oxford-result})
to the integrand of (\ref{bogoliubov outcoupling current}). Thus
our result for the out-coupling current reduces to that derived in
Refs.\( \:  \)\cite{burnett-1,burnett-2}, where the out-coupled
atoms were in free-particle states.

\section{The Local Density Approximation (LDA)}

\label{sec:LDA}In this section, we evaluate the out-coupling current
in (\ref{bogoliubov outcoupling current}) and estimate the magnitude
and temperature-dependence of the quantum evaporation and pair-breaking
effects discussed in Section \ref{sec:Case of Out-coupling to a non-Bose-condensed Gas}.
In previous work \cite{burnett-1,burnett-2}, these noncondensate
processes were found to be a few percent, relative to the dominant
contribution of atoms tunneling out of the high density condensate.
It has also been found \cite{unpublished-structure-factor-finite-T}
that in current Bose gas experiments, the contribution of the noncondensate
atoms to the dynamic structure factor \( S(\mathbf{q},\omega ) \)
at \( T=0.9T_{BEC} \) is about \( 5\, \% \) of the contribution
from the condensate fluctuations (the latter has been studied in Refs.~\cite{structure-factor-phonons-exp,structure-factor-phonons-houches-exp,trento-dynamic-structure-factor}
at \( T=0 \)). This small amplitude is, of course, a direct reflection
of the strongly peaked condensate at the centre of the trap compared
to the broad, low density profile of the noncondensate atoms. However,
while small, the tunneling current from the noncondensate atoms gives
direct information about the Bose-induced correlations and thus is
a worthwhile goal for future Bragg scattering and Raman experiments.

The condensate current in (\ref{bogoliubov outcoupling current})
can be rewritten in the form\begin{equation}
\label{condensate outcoupling current}
\delta I_{out}^{c}(\mathbf{q},\delta )\equiv \gamma ^{2}\int \frac{\mathrm{d}\mathbf{k}}{(2\pi )^{3}}\left| \Phi _{10}(\mathbf{q}-\mathbf{k})\right| ^{2}\delta (\nu _{\mathbf{k}}),
\end{equation}
 where \( \Phi _{10}(\mathbf{q}-\mathbf{k}) \) is the Fourier transform
of the Bose order parameter \begin{equation}
\label{order parameter Fourier transform definition}
\Phi _{10}(\mathbf{q}-\mathbf{k})\equiv \int \mathrm{d}\mathbf{r}\mathrm{e}^{i(\mathbf{q}-\mathbf{k})\cdot \mathbf{r}}\Phi _{10}(\mathbf{r})
\end{equation}
 and we have introduced the abbreviation \( \nu _{\mathbf{k}}\equiv k^{2}/2m-(\omega -\Delta \varepsilon )-\mu _{1TF} \).
It is also useful to introduce a {}``detuned frequency'' \begin{equation}
\label{definition of detuning parameter}
\delta \equiv \omega -\Delta \varepsilon .
\end{equation}
 The noncondensate contribution in (\ref{bogoliubov outcoupling current})
is more conveniently written in terms of the trapped gas Bogoliubov
spectral density in (\ref{bogoliubov-spectral-density-frequency}),
\begin{equation}
\label{noncondensate outcoupling current}
\delta I_{out}^{nc}(\mathbf{q},\delta )=\gamma ^{2}\int \mathrm{d}\mathbf{r}\int \mathrm{d}\mathbf{r}^{\prime }\int \frac{\mathrm{d}\mathbf{k}}{(2\pi )^{3}}\mathrm{e}^{i(\mathbf{q}-\mathbf{k})\cdot (\mathbf{r}-\mathbf{r}^{\prime })}N_{1}^{0}(\nu _{\mathbf{k}})A_{\tilde{\psi }_{1}^{\dagger }\tilde{\psi }_{1}}(\mathbf{r}^{\prime },\mathbf{r},\nu _{\mathbf{k}}).
\end{equation}
 In the rest of this section, for notational simplicity, we shall
drop the index {}``1'' on the quantities associated with the trapped
gas. Equations (\ref{condensate outcoupling current}) and (\ref{noncondensate outcoupling current})
summarize our main results for the tunneling currents into the vacuum,
showing very clearly which properties of the trapped Bose gas they
depend on.

\subsection{Condensate contribution}

In the standard Thomas-Fermi approximation, the condensate density
profile is given by \cite{rmp-bose-gas-review} \begin{equation}
\label{Thomas-Fermi condensate density}
n_{c0}(\mathbf{r})=\left\{ \begin{array}{ccc}
\biggr (\mu _{TF}-U_{ext}(\mathbf{r})\biggl )/g, & r\leq R_{TF} & \\
0, & r>R_{TF}
\end{array}\right. ,
\end{equation}
 where the chemical potential is\begin{equation}
\label{Thomas-Fermi chemical potential}
\mu _{TF}=\frac{1}{2}m\omega _{0}^{2}R_{TF}^{2},
\end{equation}
 and \begin{equation}
\label{Thomas-Fermi radius}
R_{TF}=a_{HO}\left( 15N_{c}(T)\frac{a}{a_{HO}}\right) ^{1/5}.
\end{equation}
 The Thomas-Fermi radius \( R_{TF} \) defines the size of the condensate,
with \( N_{c}(T) \) the total number of particles in the Bose-condensate
at temperature \( T \) and \( a_{HO}=\sqrt{\hbar /m\omega _{0}} \)
is the oscillator length. The \( T=0 \) Thomas-Fermi results have
been extended to finite temperature by adjusting \( N_{c} \) according
to the noninteracting trapped gas result \cite{rmp-bose-gas-review} \begin{equation}
\label{noninteracting gas depletion}
\frac{N_{c}(T)}{N}=1-\left( \frac{T}{T_{BEC}}\right) ^{3},
\end{equation}
 where \( k_{B}T_{BEC}\equiv 0.94\hbar \omega _{0}N^{1/3} \) is the
noninteracting trapped gas Bose-condensation temperature. Using (\ref{Thomas-Fermi condensate density})
for the order parameter \( \Phi _{0}(\mathbf{r})=\sqrt{n_{c0}(\mathbf{r})} \),
the Fourier transform in (\ref{order parameter Fourier transform definition})
can be calculated analytically \begin{eqnarray}
\Phi _{0}(\mathbf{k}) & = & 4\pi \int _{0}^{\infty }\mathrm{d}r\, r\sqrt{n_{c0}(r)}\frac{\sin (kr)}{k}\nonumber \\
 & = & \frac{2\pi ^{2}R_{TF}^{2}}{\sqrt{2g}k^{2}}J_{2}(kR_{TF}),\label{order parameter Fourier transform} 
\end{eqnarray}
 where \( J_{2} \) is the ordinary Bessel function of order 2. This
result is the same as previously obtained \cite{baym-momentum-condensate-tf,trento-dynamic-structure-factor},
taking into account that we use harmonic oscillator units (length
is measured in units of \( a_{HO}=\sqrt{\hbar /m\omega _{0}} \) and
energy is measured in terms of \( \hbar \omega _{0} \)).

Writing the integral in (\ref{condensate outcoupling current}) in
terms of spherical coordinates, we obtain using (\ref{order parameter Fourier transform})
\begin{eqnarray}
\delta I_{out}^{c}(\mathbf{q},\delta ) & = & \frac{\gamma ^{2}}{2\pi \omega _{0}}\frac{\big (gn_{c0}(\mathbf{r}=0)\big )^{7/2}}{g}\nonumber \\
 &  & \times \frac{1}{q}\int ^{x_{+}}_{x_{-}}\mathrm{d}x\frac{J_{2}^{2}(x)}{x^{4}}\theta (\delta +\mu _{TF}),\label{LDA condensate outcoupling contribution in harmonic oscillator units} 
\end{eqnarray}
 where \( x_{\pm }=\sqrt{2gn_{c0}(\mathbf{r}=0)}\bigg (q\pm \sqrt{2(\delta +\mu _{TF})}\bigg )^{2} \).
The result (\ref{LDA condensate outcoupling contribution in harmonic oscillator units})
may be easily generalized to asymmetric harmonic traps, having trapping
potential \begin{equation}
\label{Uext for asymmetrictrap}
U_{ext}(\mathbf{r})=\frac{1}{2}m\bigg (\omega _{x}^{2}x^{2}+\omega ^{2}_{y}y^{2}+\omega _{z}^{2}z^{2}\bigg ),
\end{equation}
 by introducing the coordinate transformation\begin{equation}
\label{coordinate transform for asymmetric trap to symmetric trap}
x=\frac{\bar{\omega }}{\omega _{x}}x^{\prime },\quad y=\frac{\bar{\omega }}{\omega _{y}}y^{\prime },\quad z=\frac{\bar{\omega }}{\omega _{z}}z^{\prime },
\end{equation}
 where \( \bar{\omega }\equiv (\omega _{x}\omega _{y}\omega _{z})^{1/3} \).
The same procedure as used above follows through with the same final
expressions, apart from \( \omega _{0} \) being replaced by \( \bar{\omega } \).
In addition, any wavevector \( \mathbf{k} \) in our tunneling expressions
such as (\ref{condensate outcoupling current}) needs to be interpreted
as \( \mathbf{k}\rightarrow \left( \frac{\bar{\omega }}{\omega _{x}}k_{x},\frac{\bar{\omega }}{\omega _{y}}k_{y},\frac{\bar{\omega }}{\omega _{z}}k_{z}\right)  \)
in an anisotropic trap \cite{trento-dynamic-structure-factor}.

The total out-coupling current from the condensate given by (\ref{condensate outcoupling current}),
as a function of the detuning parameter \( \delta  \) in (\ref{definition of detuning parameter}),
is shown by the solid lines in Figs.~\ref{fig:condensate-noncondensate out-coupling current for q=3D3D2/aho and a=3D3D5nm}
and \ref{fig:total out-coupling current for q=3D3D2/aho a=3D3D5nm T=3D3D0.2TBEC}.
In all our numerical calculations, for the two-photon Rabi coupling
we have used \( \gamma =0.2\omega _{0} \). Note the logarithmic scale
for the out-coupled atom current. The current magnitude only decreases
as a function of increasing temperature by about \( 20\, \% \). From
Fig.~\ref{fig:condensate-noncondensate out-coupling for q=3D3D10/aho and a=3D3D5nm},
it can be seen that the maximum condensate tunneling current amplitude
is sensitive to the magnitude of the momentum transfer \( \mathbf{q} \).
The decrease in the maximum tunneling current with increasing values
of \( q \) is a result of the fact that the momentum distribution
of atoms in the condensate is peaked at zero momentum, decreasing
for increasing momentum. For an asymmetric trap, this means that the
directions of strongest confinement (largest trap frequency) will
out-couple most efficiently. We observe also that the peak of the
condensate tunneling current shifts toward positive \( \delta  \)
as \( q \) increases.

\subsection{Noncondensate contribution}

The out-coupling current (\ref{noncondensate outcoupling current})
from the noncondensate atoms will be estimated using a simple local
density approximation (LDA). The LDA has also been recently used to
calculate \( S(\mathbf{q},\omega ) \) in trapped gases \cite{timmermans-tommasini-structure-factor-lda,trento-dynamic-structure-factor}
. We first introduce a new coordinate system defined in terms of the
relative coordinate\begin{equation}
\label{relativecoordinate}
\bar{\mathbf{r}}=\mathbf{r}-\mathbf{r}^{\prime }
\end{equation}
 and the center-of-mass coordinate\begin{equation}
\label{center-of-mass coordinate}
\mathbf{R}=\frac{\mathbf{r}+\mathbf{r}^{\prime }}{2}.
\end{equation}
 Writing the single-particle spectral density function \( A_{\tilde{\psi }^{\dagger }\tilde{\psi }}(\mathbf{r},\mathbf{r}^{\prime },\omega ) \)
in (\ref{noncondensate outcoupling current}) in terms of these new
coordinates, we note that for large condensates and quasiparticle
energies much larger than \( \hbar \omega _{0} \), the spectral density
will vary slowly as a function of \( \mathbf{R} \). It is thus plausible
to approximate \( A_{\tilde{\psi }^{\dagger }\tilde{\psi }} \) by
its homogeneous form, but adjusted so that at each point \( \mathbf{R} \)
the appropriate condensate density \( n_{c0}(\mathbf{R}) \) is used.
Thus, in the LDA, the single-particle spectral density function in
(\ref{noncondensate outcoupling current}) is approximated by\begin{equation}
\label{LDA approximation for spectral density}
A_{\tilde{\psi }^{\dagger }\tilde{\psi }}(\mathbf{r},\mathbf{r}^{\prime },\omega )=A_{\tilde{\psi }^{\dagger }\tilde{\psi }}(\bar{\mathbf{r}},\mathbf{R},\omega )\simeq A_{\tilde{\psi }^{\dagger }\tilde{\psi }}^{LDA}(\bar{\mathbf{r}},\omega ;n_{c0}(\mathbf{R})),
\end{equation}
 where \( A_{\tilde{\psi }^{\dagger }\tilde{\psi }}^{LDA} \) denotes
the \emph{uniform} gas expression for the spectral density function
for a condensate density \( n_{c0} \).

In the Bogoliubov-Popov approximation, the LDA form (\ref{LDA approximation for spectral density})
for the spectral density in energy-momentum space is \cite{giorgini-popov-and-semiclassical}\begin{eqnarray}
A_{\tilde{\psi }^{\dagger }\tilde{\psi }}^{LDA}(\mathbf{p},\omega ;n_{c0}(\mathbf{R})) & = & 2\pi \Biggl [u^{2}(\mathbf{p};n_{c0}(\mathbf{R}))\delta \biggl (\omega -E(\mathbf{p};n_{c0}(\mathbf{R}))\biggr )\nonumber \\
 &  & -v^{2}(\mathbf{p};n_{c0}(\mathbf{R}))\delta \biggl (\omega +E(\mathbf{p};n_{c0}(\mathbf{R}))\biggr )\Biggr ].\label{LDA Bogoliubov spectral density in momentum space} 
\end{eqnarray}
 Thus, in the LDA the quasiparticle energies become \begin{equation}
\label{LDA quasiparticle energy}
E(\mathbf{p};n_{c0}(\mathbf{R}))=\left\{ \begin{array}{ccc}
\sqrt{\varepsilon _{\mathbf{p}}^{2}+2\mu _{eff}(\mathbf{R})\varepsilon _{\mathbf{p}}}, & R\leq R_{TF} & \\
\varepsilon _{\mathbf{p}}-\mu _{eff}(\mathbf{R}), & R>R_{TF}
\end{array}\right. ,
\end{equation}
 while the Bogoliubov quasiparticle mode amplitudes are given by\begin{eqnarray}
u^{2}(\mathbf{p};n_{c0}(\mathbf{R})) & = & \left\{ \begin{array}{ccc}
\frac{\varepsilon _{\mathbf{p}}-\mu _{eff}(\mathbf{R})}{2E(\mathbf{p};n_{c0}(\mathbf{R}))}+\frac{1}{2}, & R\leq R_{TF} & \\
1, & R>R_{TF}
\end{array}\right. \label{LDA quasiparticle u mode} \\
v^{2}(\mathbf{p};n_{c0}(\mathbf{R})) & = & \left\{ \begin{array}{ccc}
\frac{\varepsilon _{\mathbf{p}}-\mu _{eff}(\mathbf{R})}{2E(\mathbf{p};n_{c0}(\mathbf{R}))}-\frac{1}{2}, & R\leq R_{TF} & \\
0, & R>R_{TF}
\end{array}\right. .\label{LDA quasiparticle v mode} 
\end{eqnarray}
 We have introduced the free atom energy \( \varepsilon _{\mathbf{p}}\equiv p^{2}/2m \)
and\begin{equation}
\label{effective chemical potential}
\mu _{eff}(\mathbf{R})\equiv \mu _{TF}-U_{ext}(\mathbf{R})-2gn_{c0}(\mathbf{R}),
\end{equation}
 plays the role of an {}``effective\char`\"{} local chemical potential,
modifying the chemical potential \( \mu _{TF} \) as a result of the
inhomogeneity of the trap.

Within the LDA, the noncondensate contribution (\ref{noncondensate outcoupling current})
to the out-coupling current is given by\begin{equation}
\label{LDA noncondensate outcoupling current version 1}
\delta I_{out}^{nc}(\mathbf{q},\delta )=\frac{\gamma ^{2}}{\omega _{0}}\int \frac{\mathrm{d}\mathbf{k}}{(2\pi )^{3}}N^{0}(\nu _{\mathbf{k}})\int \mathrm{d}\mathbf{R}A^{LDA}_{\tilde{\psi }^{\dagger }\tilde{\psi }}(\mathbf{q}-\mathbf{k},\nu _{\mathbf{k}};n_{c0}(\mathbf{R})),
\end{equation}
 where as before we have introduced the variable \( \nu _{\mathbf{k}}\equiv \varepsilon _{\mathbf{k}}-\delta -\mu _{TF} \)
for notational simplicity. Using the identity\begin{equation}
\label{delta function identity}
\delta (f(x))=\sum _{i}\frac{1}{\left| f^{\prime }(x_{i})\right| }\delta (x-x_{i}),
\end{equation}
 where \( f(x_{i})=0 \), we may evaluate the integral over \( \mathbf{R} \)
in (\ref{LDA noncondensate outcoupling current version 1}) assuming
a spherically symmetric trap. After some algebra, one obtains the
following expression for the noncondensate contribution to the out-coupling
current\begin{equation}
\label{LDA noncondensate outcoupling current in harmonic oscillator units}
\delta I_{out}^{nc}(\mathbf{q},\delta )=\frac{2\pi \gamma ^{2}}{\omega _{0}}\int \frac{\mathrm{d}\mathbf{k}}{(2\pi )^{3}}N^{0}(\nu _{\mathbf{k}})\bar{A}^{LDA}_{\tilde{\psi }^{\dagger }\tilde{\psi }}(\mathbf{q}-\mathbf{k},\nu _{\mathbf{k}}),
\end{equation}
 where we have introduced the LDA spatially averaged spectral density
function in energy-momentum space\begin{eqnarray}
\bar{A}^{LDA}_{\tilde{\psi }^{\dagger }\tilde{\psi }}(\mathbf{k},\omega ) & \equiv  & \int \mathrm{d}\mathbf{R}A^{LDA}_{\tilde{\psi }^{\dagger }\tilde{\psi }}(\mathbf{k},\omega ;n_{c0}(\mathbf{R}))\nonumber \\
 & = & 2\pi \sqrt{2\bigg (\mu _{TF}-\mu (\mathbf{k},\omega )\bigg )}\left( \frac{\varepsilon _{\mathbf{k}}^{2}+\omega ^{2}}{2\varepsilon _{\mathbf{k}}^{2}}+\frac{\left| \omega \right| }{\varepsilon _{\mathbf{k}}}\right) \theta (\omega )\theta \Biggl (\mu (\mathbf{k},\omega )\biggl (\mu _{TF}-\mu (\mathbf{k},\omega )\biggr )\Biggr )\nonumber \\
 &  & +2\pi \sqrt{2\bigg (\mu _{TF}-\mu (\mathbf{k},\omega )\bigg )}\left( -\frac{\varepsilon _{\mathbf{k}}^{2}+\omega ^{2}}{2\varepsilon _{\mathbf{k}}^{2}}+\frac{\left| \omega \right| }{\varepsilon _{\mathbf{k}}}\right) \theta (-\omega )\theta \Biggl (\mu (\mathbf{k},\omega )\biggl (\mu _{TF}-\mu (\mathbf{k},\omega )\biggr )\Biggr )\nonumber \\
 &  & +4\pi \sqrt{2(\omega -\varepsilon _{\mathbf{k}}+\mu _{TF})}\theta (\omega -\varepsilon _{\mathbf{k}}),\label{LDA energy-momentum space spectral density function} 
\end{eqnarray}
 where\begin{equation}
\label{mukw}
\mu (\mathbf{k},\omega )\equiv \frac{\omega ^{2}-\varepsilon _{\mathbf{k}}^{2}}{2\varepsilon _{\mathbf{k}}}
\end{equation}
 and harmonic oscillator units are used. Note that the noncondensate
out-coupling current in (\ref{LDA noncondensate outcoupling current in harmonic oscillator units})
may be generalized to an asymmetric trap in the same manner as discussed
earlier for the condensate out-coupling current (see remarks below
(\ref{coordinate transform for asymmetric trap to symmetric trap})).

The formula in (\ref{LDA noncondensate outcoupling current in harmonic oscillator units})
has a simple physical interpretation. The integrand gives the number
of emitted atoms with momentum \( \mathbf{k} \) per second. The first
term in (\ref{LDA energy-momentum space spectral density function})
is associated with the positive energy pole in (\ref{bogoliubov-spectral-density-frequency}),
with amplitude \( u^{2} \), and only occurs for positive frequency
\( \omega >0 \). The second term in (\ref{LDA energy-momentum space spectral density function})
arises from the negative energy pole in (\ref{bogoliubov-spectral-density-frequency}),
with amplitude \( v^{2} \), and only occurs for negative frequency
\( \omega <0 \). These first two terms are associated with quasiparticles
in the region of the (Thomas-Fermi) condensate, i.e.~for \( R\leq R_{TF} \).
The third term, in contrast, is associated with quasiparticles lying
outside of the condensate region. These quasiparticles comprise a
Hartree-Fock gas in our simple model calculation. The dominant weight
in (\ref{LDA energy-momentum space spectral density function}) comes
from small \( \varepsilon _{\mathbf{k}} \) and \( \omega  \). The
quasiparticles with \( u^{2} \) and \( v^{2} \) weights in the region
where there is a condensate are the dominant contribution. For \( \omega >\varepsilon _{\mathbf{k}}(>0) \),
one also finds a contribution from the normal Hartree-Fock gas of
atoms existing outside of the condensate. As the temperature increases
toward \( T_{BEC} \), this contribution from the normal Hartree-Fock
gas becomes more significant relative to the Bogoliubov quasiparticles.
Of course, for \( T>T_{BEC} \), it is the sole contribution.

\begin{figure}[p]
{\centering \includegraphics{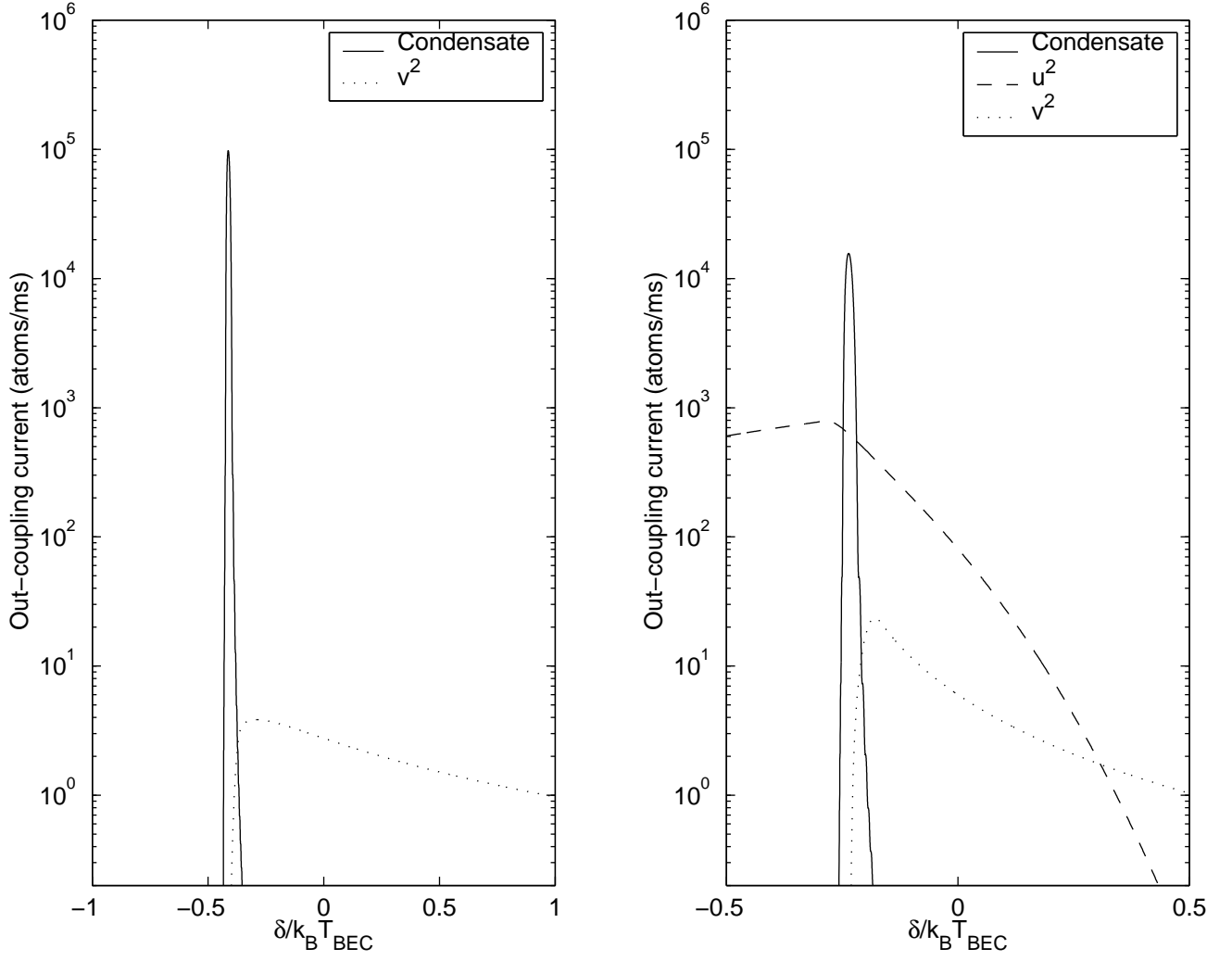} \par}

\caption{\label{fig:condensate-noncondensate out-coupling current for q=3D3D2/aho and a=3D3D5nm}Out-coupling
atom current for \protect\( a\simeq 5\, nm\protect \) and \protect\( q=2/a_{HO}\protect \).
The detuned frequency \protect\( \delta \protect \) is defined in
(\ref{definition of detuning parameter}). The plot on the left is
at \protect\( T=0\protect \). The plot on the right is at \protect\( T=0.9T_{BEC}\protect \).
In Figs.~\ref{fig:condensate-noncondensate out-coupling current for q=3D3D2/aho and a=3D3D5nm}-\ref{fig:total out-coupling current for q=3D3D10/aho a=3D3D5nm T=3D3D0.2TBEC},
we take \protect\( N=10^{6}\protect \) atoms, \protect\( \omega _{0}=2\pi \times 60\, Hz\protect \)
and \protect\( \gamma =0.2\omega _{0}\protect \).}
\end{figure}

\begin{figure}[p]
{\centering \includegraphics{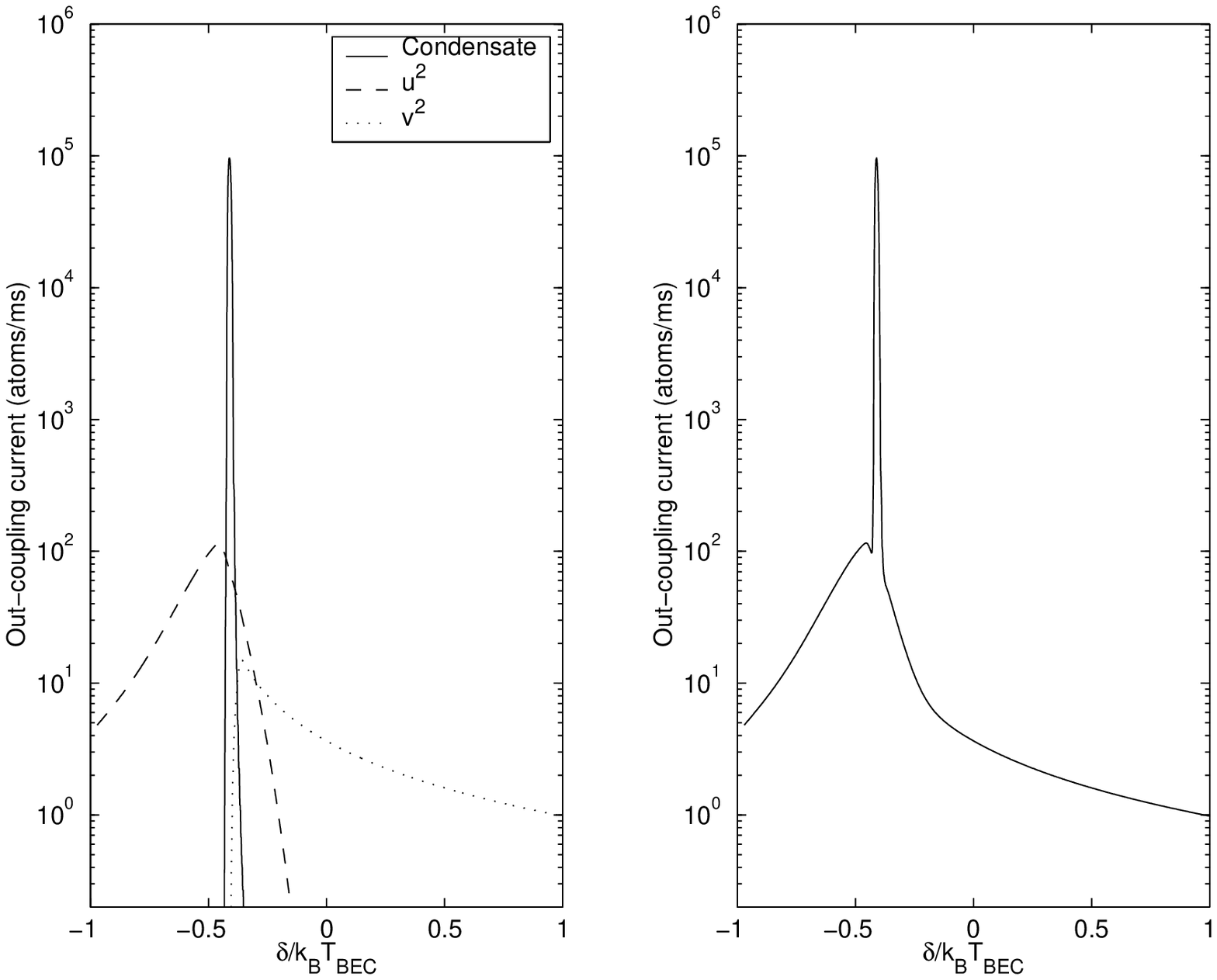} \par}

\caption{\label{fig:total out-coupling current for q=3D3D2/aho a=3D3D5nm T=3D3D0.2TBEC}Out-coupling
atom current for \protect\( a\simeq 5\, nm\protect \) and \protect\( q=2/a_{HO}\protect \).
The plot on the left shows each component contributing to the out-coupling
atom current. The plot on the right shows the total out-coupling atom
current. In both plots \protect\( T=0.2T_{BEC}\protect \).}
\end{figure}

We have evaluated the integral appearing in (\ref{LDA noncondensate outcoupling current in harmonic oscillator units})
numerically. Some results, as a function of the detuning parameter
\( \delta  \) for various temperatures, are shown in Figs.~\ref{fig:condensate-noncondensate out-coupling current for q=3D3D2/aho and a=3D3D5nm}
and \ref{fig:total out-coupling current for q=3D3D2/aho a=3D3D5nm T=3D3D0.2TBEC}.
Note the logarithmic scale for the out-coupled atom current. As a
function of detuning, the out-coupling current separates into two
regions. From (\ref{LDA noncondensate outcoupling current in harmonic oscillator units})
and (\ref{LDA energy-momentum space spectral density function}),
we observe that for \( \delta <-\mu _{TF} \) (or \( \omega +\varepsilon _{1}+\mu _{TF}<\varepsilon _{2} \)),
we have \( \nu _{\mathbf{k}}>0 \) for all values of the wavevector
\( \mathbf{k} \). In this region, we find a contribution from the
\( u^{2} \) Bogoliubov modes, in addition to a smaller contribution
from the atoms coming from outside the condensate region, at finite
temperatures. This contribution to the out-coupling current \( \delta I_{N}^{nc} \)
represents the {}``quantum evaporation'' process discussed at the
end of Section \ref{sec:Case of Out-coupling to a non-Bose-condensed Gas},
being associated with positive energy poles of the single-particle
noncondensate Green's function. It is only present at finite temperatures.
In Fig.~\ref{fig:total out-coupling current for q=3D3D2/aho a=3D3D5nm T=3D3D0.2TBEC},
we have plotted the \emph{total} out-coupling current as a function
of the detuning \( \delta  \) at an intermediate temperature of \( T=0.2T_{BEC} \).

\begin{figure}[p]
{\centering \includegraphics{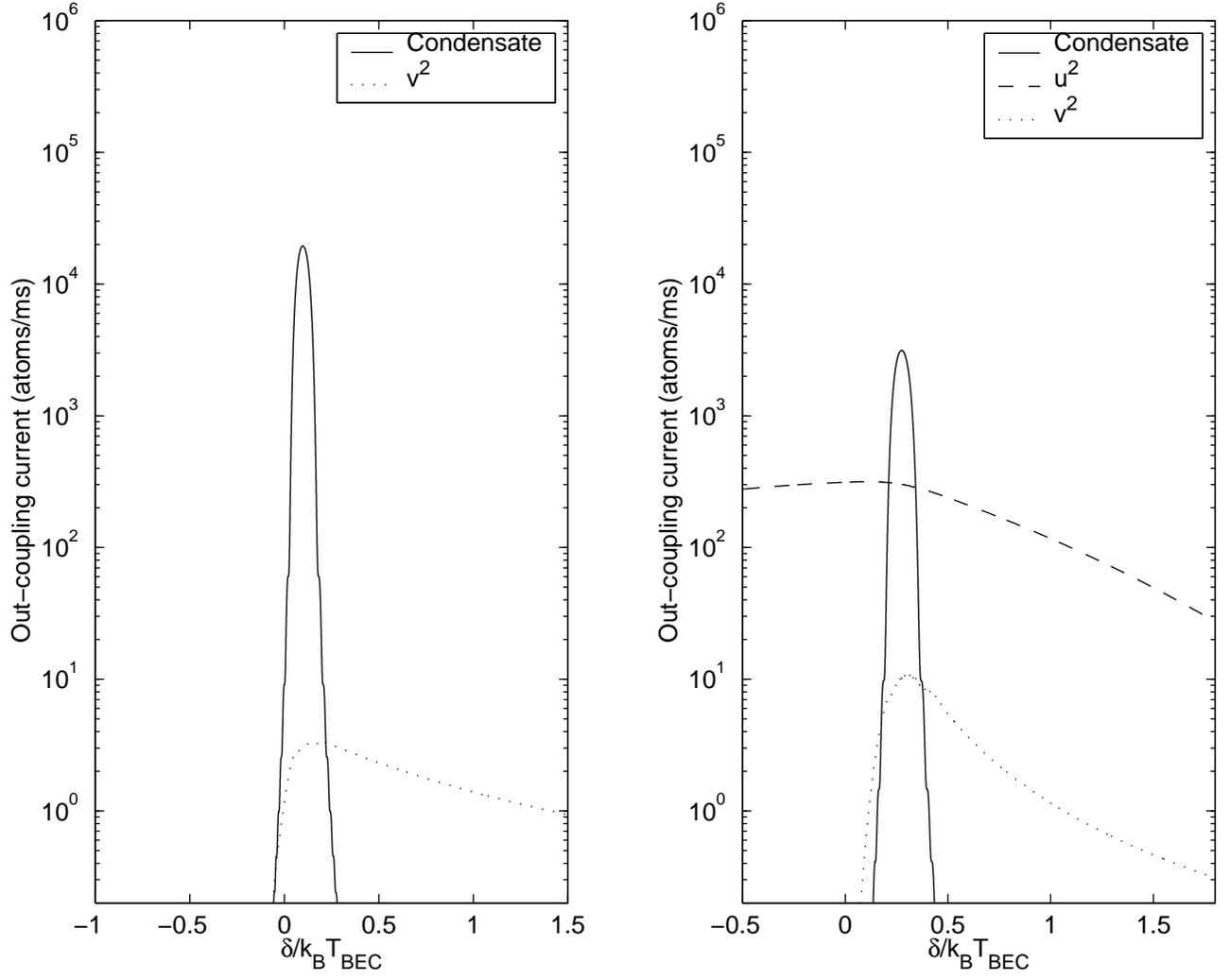} \par}

\caption{\label{fig:condensate-noncondensate out-coupling for q=3D3D10/aho and a=3D3D5nm}Out-coupling
current for \protect\( a\simeq 5\, nm\protect \) and \protect\( q=10/a_{HO}\protect \).
The plot on the left is at \protect\( T=0\protect \). The plot on
the right is at \protect\( T=0.9T_{BEC}\protect \).}
\end{figure}

\begin{figure}[p]
{\centering \includegraphics{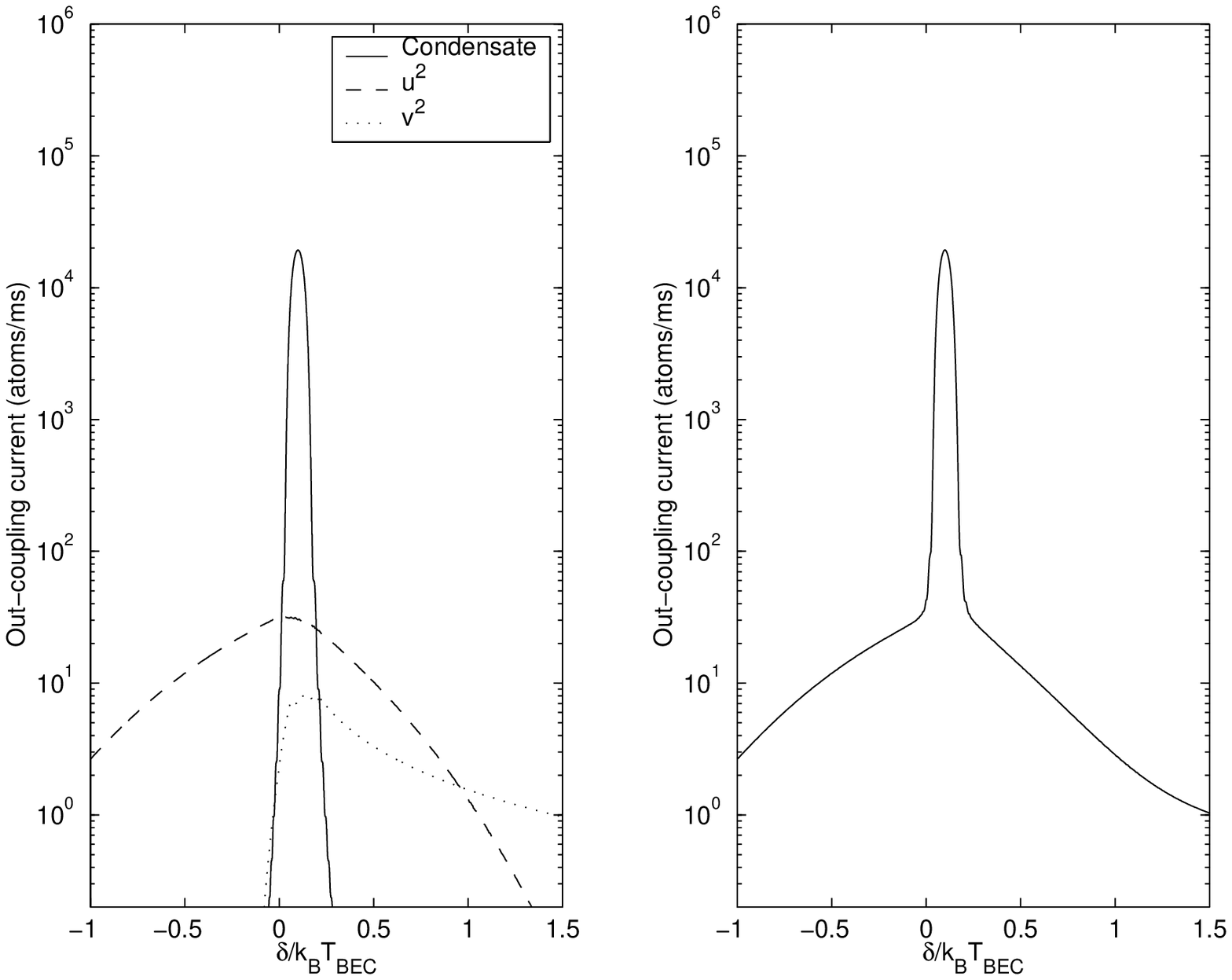} \par}

\caption{\label{fig:total out-coupling current for q=3D3D10/aho a=3D3D5nm T=3D3D0.2TBEC}Out-coupling
atom current for \protect\( a\simeq 5\, nm\protect \) and \protect\( q=10/a_{HO}\protect \).
The plot on the left shows each component contributing to the out-coupling
atom current. The plot on the right shows the total out-coupling atom
current. In both plots \protect\( T=0.2T_{BEC}\protect \).}
\end{figure}

However, for \( \delta >-\mu _{TF} \) (or \( \omega +\varepsilon _{1}+\mu _{TF}>\varepsilon _{2} \)),
\( \nu _{\mathbf{k}} \) can be less than zero for a restricted set
of wavevectors \( \mathbf{k} \). In this case, there are two contributions
to the out-coupling current \( \delta I_{N}^{nc} \). One arises from
the \( v^{2} \) Bogoliubov modes and hence represents the so-called
pair-breaking process associated with the negative energy poles of
the single-particle noncondensate spectral density function (i.e.,~a
tunneling atom creates a Bogoliubov excitation in the trapped gas).
This is the only process which contributes to the noncondensate out-coupling
current at zero temperature. At finite temperature, a second process
arises which is associated with quantum evaporation from high momentum
states. This involves the positive energy poles of the single-particle
noncondensate spectral density function. This second contribution
decays more quickly as the detuning is increased, compared with the
\( v^{2} \) contribution. We note that these two regions will be
Doppler shifted toward positive \( \delta  \) by the nonzero momentum
kick \( \mathbf{q} \) from the laser fields. The overall behavior
of our LDA results in Figs.~\ref{fig:condensate-noncondensate out-coupling current for q=3D3D2/aho and a=3D3D5nm}-\ref{fig:total out-coupling current for q=3D3D10/aho a=3D3D5nm T=3D3D0.2TBEC}
is similar to that reported in Refs.~\cite{burnett-1,burnett-2},
which were based on solving a one-dimensional Gross-Pitaevskii equation
for an isotropic trap.

We have also evaluated the noncondensate contribution to the out-coupling
current for \( q=10/a_{HO} \), with the results shown in Figs.~\ref{fig:condensate-noncondensate out-coupling for q=3D3D10/aho and a=3D3D5nm}
and \ref{fig:total out-coupling current for q=3D3D10/aho a=3D3D5nm T=3D3D0.2TBEC}.
We observe that the point at which there is no longer any contribution
from the \( v^{2} \) Bogoliubov modes shifts increasingly toward
positive \( \delta  \) as \( q \) is increased. However, unlike
for the case of \( q=2/a_{HO} \), the contribution at finite temperatures
from the \( u^{2} \) Bogoliubov modes dominates that from the \( v^{2} \)
Bogoliubov modes for the range of \( \delta  \) considered. This
results from the fact that the higher momentum quasiparticles have
greater spectral weight. We have plotted the \emph{total} out-coupling
current at a temperature of \( T=0.2T_{BEC} \) in Fig.~\ref{fig:total out-coupling current for q=3D3D10/aho a=3D3D5nm T=3D3D0.2TBEC}.

Because of the strength of the condensate contribution, some comment
is required about the observability of the noncondensate contribution
in the total out-coupling current. Since the contribution from the
condensate is smallest at high \( q \), we compare the condensate
and noncondensate contributions at a momentum kick of \( q=10/a_{HO} \)
(see Fig.~\ref{fig:condensate-noncondensate out-coupling for q=3D3D10/aho and a=3D3D5nm}).
Since the out-coupling current as a function of \( \delta  \) is
much broader for the noncondensate contribution than the condensate
contribution, there are regions in which the strong contribution from
the condensate will not dwarf that from the noncondensate. At \( T=0.9T_{BEC} \),
the maximum contribution from the quantum evaporation process is about
\( 10\, \% \) of the maximum contribution from the condensate (see
Fig.~\ref{fig:condensate-noncondensate out-coupling for q=3D3D10/aho and a=3D3D5nm}).
The small but interesting \( v^{2} \) Bogoliubov contribution is
perhaps more visible as a {}``distinct contribution'' in the limit
of very low temperatures (see the \( T=0 \) results in Figs.~\ref{fig:condensate-noncondensate out-coupling current for q=3D3D2/aho and a=3D3D5nm}
and \ref{fig:condensate-noncondensate out-coupling for q=3D3D10/aho and a=3D3D5nm}).

The unique features associated with the out-coupled atoms coming from
the noncondensate can also be enhanced by working with larger values
of the s-wave scattering length \( a \). In particular, one can {}``tune''
the scattering length \( a \) significantly by the use of Feshbach
resonances \cite{stoof-feshbach-resonance-2,ketterle-feshbach-1}.
The amplification of the out-coupling current of atoms associated
with the negative energy poles of the single-particle noncondensate
Green's function could be looked for in trapped gases near a Feshbach
resonance. Out-coupling experiments may thus provide important information
about the Bose-condensed gas excitation spectrum if we can increase
the s-wave scattering length substantially. In this case, of course,
we expect observable changes from the simple Bogoliubov-Popov approximation
we introduced in Section \ref{Section: Coupling between two trapped gases}
for the single-particle spectral density. Such many-body effects can
thus be probed {}``directly'' using the out-coupling kind of experiment
we have been discussing.

\section{Conclusions}

This paper has studied the weak coupling of two trapped dilute Bose
gases using a tunneling Hamiltonian approach. When considering tunneling
between two trapped condensates, we obtained a coherent Josephson-type
time-dependent contribution given in (\ref{equilibrium current 3}),
with a frequency \( \Omega \equiv \omega -\Delta \varepsilon -\Delta \mu  \)
(see also Ref.~\cite{finite-T-josephson}). The expression in (\ref{equilibrium current 3}),
which corresponds to coherent transfer of atoms between two condensates,
is first order in the coupling \( \gamma  \). In addition, we found
that two types of quasiparticle currents emerge from treating a perturbation
of the kind shown in (\ref{V}) to second order: a static, time-independent
current given by (\ref{deltaINa}) and (\ref{short time normal current})
involving only regular correlation functions of the form \( \left\langle \widetilde{\psi }^{\dagger }\widetilde{\psi }\right\rangle  \),
and a time-dependent current given by (\ref{deltaIAa}) and (\ref{deltaIAb})
involving a second harmonic with frequency \( 2\Omega  \). The latter
is associated with anomalous correlation functions of the kind \( \left\langle \widetilde{\psi }\widetilde{\psi }\right\rangle  \),
and is reminiscent of the Josephson current carried by Cooper pairs
in superconductors \cite{mahan}. These tunneling currents were written
in terms of single-particle spectral density functions that contain
all the details of the microscopic correlations of the Bose-condensed
gases. In this manner, the tunneling dynamics is cleanly \emph{separated}
from the specific approximation used for the many-body description
of the Bose-condensed gas. This separation was one of the main goals
of our work. 

Our paper further serves to unify and extend the work of Ref.~\cite{finite-T-josephson}
on Josephson-type and other kinds of tunneling phenomena in trapped
Bose gases, as well as recent work \cite{burnett-1,burnett-2} dealing
with out-coupling from a single Bose-condensed gas. It provides a
transparent way of expressing the various contributions to the tunneling
processes involving an inhomogeneous trapped Bose gas, before any
specific approximation is introduced. The present paper thus provides
a foundation upon which one can study the details of different tunneling
processes within various many-body approximations for the dynamics
of a Bose gas. Our work re-emphasizes the potential usefulness of
the kind of out-coupling experiments discussed in Refs.\( \:  \)\cite{burnett-1,burnett-2}
as a probe of the many-body physics of a Bose-condensed gas, including
systems of reduced dimensionality.

As we have noted in the text, our present study contrasts with the
work reported in Refs.~\cite{structure-factor-phonons-exp,structure-factor-phonons-houches-exp,timmermans-tommasini-structure-factor-lda,trento-dynamic-structure-factor,otago-bragg-structure-factor}.
The latter deal with the density fluctuation spectrum while our present
paper (see also Ref.~\cite{burnett-1,burnett-2}) probes the single-particle
Green's function of a trapped Bose gas, which is perhaps a more basic
measure of the condensate-induced correlations between atoms.

For illustration, we have used our formalism in Sections \ref{Section: Coupling between two trapped gases}
and \ref{sec:Case of Out-coupling to a non-Bose-condensed Gas} to
find a general expression for the out-coupling atom current using
the Bogoliubov-Popov quasiparticle approximation. Our results exhibit
the same three processes obtained by the Oxford group \cite{burnett-1,burnett-2}.
The first process is the tunneling of an atom out of the condensate
of the trapped gas. The second process is the {}``quantum evaporation''
of an atom from the trapped gas, \emph{destroying} a quasiparticle
excitation in the trapped gas and \emph{creating} an atom in the out-coupled
gas. The third process, referred to as a {}``pair-breaking'' process
in Ref.~\cite{burnett-1}, involves an atom tunneling out of the
trapped gas concurrently with the \emph{creation} of a quasiparticle
excitation in the trapped gas (see also Ref.~\cite{svg}). 

In Section \ref{sec:LDA}, we have numerically evaluated the relative
contributions of these three processes as a function of the laser
detuning parameter \( \delta =\omega -\Delta \varepsilon =\omega _{a}-\omega _{b}-\Delta \varepsilon  \),
for various values of the wavevector difference between the two laser
beams \( \mathbf{q}=\mathbf{k}_{a}-\mathbf{k}_{b} \) and the temperature
\( T \). This kind of experiment would be especially promising as
a probe of the correlations in a Bose gas in the case of a large scattering
length \( a \) made possible by working near a Feshbach resonance.
This enhances the tunneling contribution from the noncondensate atoms
and moreover requires the use of an improved single-particle correlation
function which includes quantum depletion of the condensate (such
as the second order Beliaev approximation).

The tunneling currents from the noncondensate atoms are admittedly
small and will require high precision experiments. We argue that the
unique information that can be obtained justifies the effort. In future
work, we plan on using the formalism set up in this paper to discuss
the case when the trapped Bose condensate is in a vortex state. In
addition, information about the single-particle correlation functions
in a one dimensional Bose gas \cite{shlyapnikov-1d} is already present
in the out-coupled tunneling current.

\begin{acknowledgments}
We would like to thank J.E.~Williams for discussions of Josephson
tunneling in trapped Bose gases, as well as M.~Imamovic-Tomasovic
and T.~Nikuni for discussions concerning the LDA. The research was
supported by NSERC of Canada.
\end{acknowledgments}

\end{document}